\pdfoutput=1
\documentclass[pra,aps,twocolumn,notitlepage,superscriptaddress,10pt]{revtex4-1}
\usepackage{graphicx,graphicx,amsthm,amsmath,amssymb}
\usepackage{wasysym}
\usepackage{color}
\usepackage{bbold}
\usepackage[caption=false]{subfig}
\usepackage[colorlinks]{hyperref}
\usepackage{float}

\newcommand{\specialcell}[2][c]{%
  \begin{tabular}[#1]{@{}c@{}}#2\end{tabular}}

\newcommand{\eat}[1]{}

\def\ket#1{| #1 \rangle}

\def\um{{\underline{\bf m}}}
\def\uq{{\underline{\bf q}}}

\def\({\left(}
\def\){\right)}

\newcommand{\googsb}{\affiliation{Google Inc., Santa Barbara, CA 93117, USA}}
\newcommand{\googla}{\affiliation{Google Inc., Venice, CA 90291, USA}}

\begin{document}

\title{What is the Computational Value of Finite Range Tunneling?}

\author{Vasil S. Denchev}
\googla
\author{Sergio Boixo}
\googla
\author{Sergei V. Isakov}
\googla
\author{Nan Ding}
\googla
\author{Ryan Babbush}
\googla
\author{Vadim Smelyanskiy}
\googla
\author{John Martinis}
\googsb
\author{Hartmut Neven}
\googla

\date{\today}

\begin{abstract}
Quantum annealing (QA) has been proposed as a quantum enhanced optimization heuristic exploiting tunneling. Here, we demonstrate how finite range tunneling can provide considerable computational advantage. For a crafted problem designed to have tall and narrow energy barriers separating local minima, the D-Wave 2X quantum annealer achieves significant runtime advantages relative to Simulated Annealing (SA). For instances with 945 variables, this results in a time-to-99\%-success-probability that is $\sim 10^8$ times faster than SA running on a single processor core. We also compared physical QA with Quantum Monte Carlo (QMC), an algorithm that emulates quantum tunneling on classical processors. We observe a substantial constant overhead against physical QA: D-Wave 2X again runs up to $\sim 10^8$ times faster than an optimized implementation of QMC on a single core. 
We note that there exist heuristic classical algorithms that can solve most instances of Chimera structured problems in a timescale comparable to the D-Wave 2X. However, we believe that such solvers will become ineffective for the next generation of annealers currently being designed. 
To investigate whether finite range tunneling will also confer an advantage for problems of practical interest, we conduct numerical studies on binary optimization problems that cannot yet be represented on quantum hardware. For random instances of the number partitioning problem, we find numerically that QMC, as well as other algorithms designed to simulate QA, scale better than SA. We discuss the implications of these findings for the design of next generation quantum annealers.
\end{abstract}

\maketitle

\section{Introduction}

Simulated annealing (SA)~\cite{kirkpatrick_optimization_1983} is perhaps the most widely used algorithm for global optimization of pseudo-Boolean
functions with little known structure. The objective function for this general class of problems is
\begin{align}
\label{eq:problem}
H_{\rm P}^{\rm cl} ({\bf s}) = -\sum_{k=1}^K \sum_{j_1\ldots j_k = 1}^N J_{j_1\cdots  j_k} s_{j_1} \cdots s_{j_k}\;,
\end{align}
where $N$ is the problem size, $s_{j} = \pm 1$ are spin
variables and the couplings $J_{j_1\ldots j_k}$ are real scalars. 
In the physics literature $H_{\rm P}^{\rm cl} ({\bf s})$ is known as the
Hamiltonian of a $K$-spin model. SA is a Monte Carlo algorithm
designed to mimic the thermalization dynamics of a system in contact
with a slowly cooling reservoir. When the temperature
is high, SA induces thermal excitations which can allow the system
to escape from local minima. As the temperature decreases, SA drives the system towards
nearby low energy spin configurations.

Almost two decades ago, quantum annealing (QA)~\cite{kadowaki_quantum_1998,*brooke_quantum_1999,*lee_global_2000,*farhi_quantum_2001,*santoro_theory_2002}
was proposed as a heuristic technique for quantum enhanced
optimization. Despite substantial academic and industrial interest~\cite{das2005quantum,das2008colloquium,harris_experimental_2010b,johnson2011quantum,bapst2013quantum,boixo2013experimental,dickson2013thermally,mcgeoch2013experimental,dash2013note,boixo_evidence_2014,lanting2014entanglement,santra2014max,ronnow_defining_2014,vinci2014hearing,shin2014quantum,vinci2014distinguishing,mcgeoch2014adiabatic,venturelli2014quantum,albash2014reexamining,king2014algorithm,crowley2014quantum,vinci2015quantum,hen2015probing,steiger2015heavy,venturelli2015quantum,bauer2015entanglement,albash2015reexamination,katzgraber2015seeking,chancellor2015maximum,perdomo2015performance,perdomo2015determination,vinci2015nested}, a unified understanding of the physics of quantum annealing and its potential as an optimization algorithm remains elusive. 
The appeal of QA relative to SA is due to the observation that quantum mechanics allows
for an additional escape route from local minima. While SA must climb over energy barriers to escape false traps, QA can penetrate these barriers without any increase in energy. This effect is a hallmark of quantum mechanics, known as quantum tunneling. The standard time-dependent Hamiltonian used for QA is
\begin{align}
  H(t) = - A(t) \sum_{j=1}^N \sigma_j^x + B(t) H_{\rm P}\;,
  \label{eq:H}
\end{align}
where $H_{\rm P}$ is written as in Eq.~\eqref{eq:problem} but with the spin variables $s_j$ replaced with $\sigma_j^z$ Pauli matrices acting on qubit $j$, and the functions $A(t)$ and $B(t)$ define the annealing schedule parameterized in terms of time $t \in [0,T_{\rm QA}]$ (see Fig.~\ref{fig:AB}). These annealing schedules can be defined in many different ways as long as the functions are smooth, $A(0) \ll B(0)$ and
$A(T_{\rm QA}) \gg B(T_{\rm QA})$. At the beginning of the annealing, the transverse field magnitude $A(t)$ is large, and the system dynamics are dominated by quantum fluctuations due to tunneling (as opposed to the thermal fluctuations used in SA).

The question of whether D-Wave processors realize computationally relevant quantum tunneling has been a subject of substantial debate. This debate has now been settled in the affirmative with a sequence of
publications~\cite{johnson2011quantum,boixo2013experimental,dickson2013thermally,boixo_evidence_2014,lanting2014entanglement,vinci2014distinguishing,albash2014reexamining} demonstrating that quantum
resources are present and functional in the processors. In particular,
Refs.~\cite{boixo_computational_2014,Boixo2016} studied the performance of the D-Wave device on problems where eight \footnote{In still unpublished experiments we saw evidence of tunneling events involving up to 12 qubits. However it is difficult to exclude higher order tunneling processes that may break up the cotunneling group.} qubit cotunneling events were employed in a functional manner to reach low-lying energy solutions.

In order to investigate the computational value of finite range tunneling in QA, we study the scaling of the exponential dependence of annealing time with the size of the tunneling domain $D$, 
\begin{equation}
T_{\rm QA}=B_{\rm QA} e^{ \alpha D}\;,
\end{equation}
where $\alpha = a_{\min} / \hbar$ and $a_{\min}$ is the rescaled
instanton action
(see Eqs.~\eqref{eq:W} and~\eqref{eq:mean_wkb}).
In SA, the system escapes from a local minimum via thermal fluctuations
over an energy barrier $\Delta E$ separating the minima. The time required for such events scales as
\begin{equation}
T_{\rm SA}=B_{\rm SA}\; e^{{\frac{ \Delta E}{k_B T}} },
\end{equation}
which is exponentially long with respect to $\Delta E$. However, for sufficiently tall and narrow barriers such that 
\begin{equation}
\frac{\Delta E}{k_B T}  >  {\alpha D} \hbox{ ,}
\end{equation}
QA can overcome barriers exponentially faster than SA. This situation was studied in Ref.~\cite{KostyaVadim2015} and it also occurs in the benchmark problems studied in this paper. 

Path integral Quantum Monte Carlo (QMC) is a method for sampling
the quantum Boltzmann distribution of a $d$ dimensional stoquastic Hamiltonian as
a marginalization of a classical Boltzmann distribution of an associated $d+1$ dimensional 
Hamiltonian. For specific cases, it was recently shown that the exponent
$\alpha$ for physical tunneling is identical to the corresponding quantity for
QMC~\cite{isakov2015understanding}. However, in the present work we find a
very substantial computational overhead associated with the prefactor
$B_{\rm QMC}$ in the expression for the runtime of QMC, $T_{\rm QMC}=B_{\rm QMC} e^{\alpha D}$. In other words, $B_{\rm QMC}$ can exceed
$B_{\rm QA}$ by many orders of magnitude.  The role of this prefactor
becomes essential in the situations where the number of cotunneling
qubits $D$ is finite, i.e., is independent of the problem size $N$ (or
depends on $N$ very weakly).

\begin{figure}[h]
  \centering
  {\includegraphics[width=0.8\columnwidth]{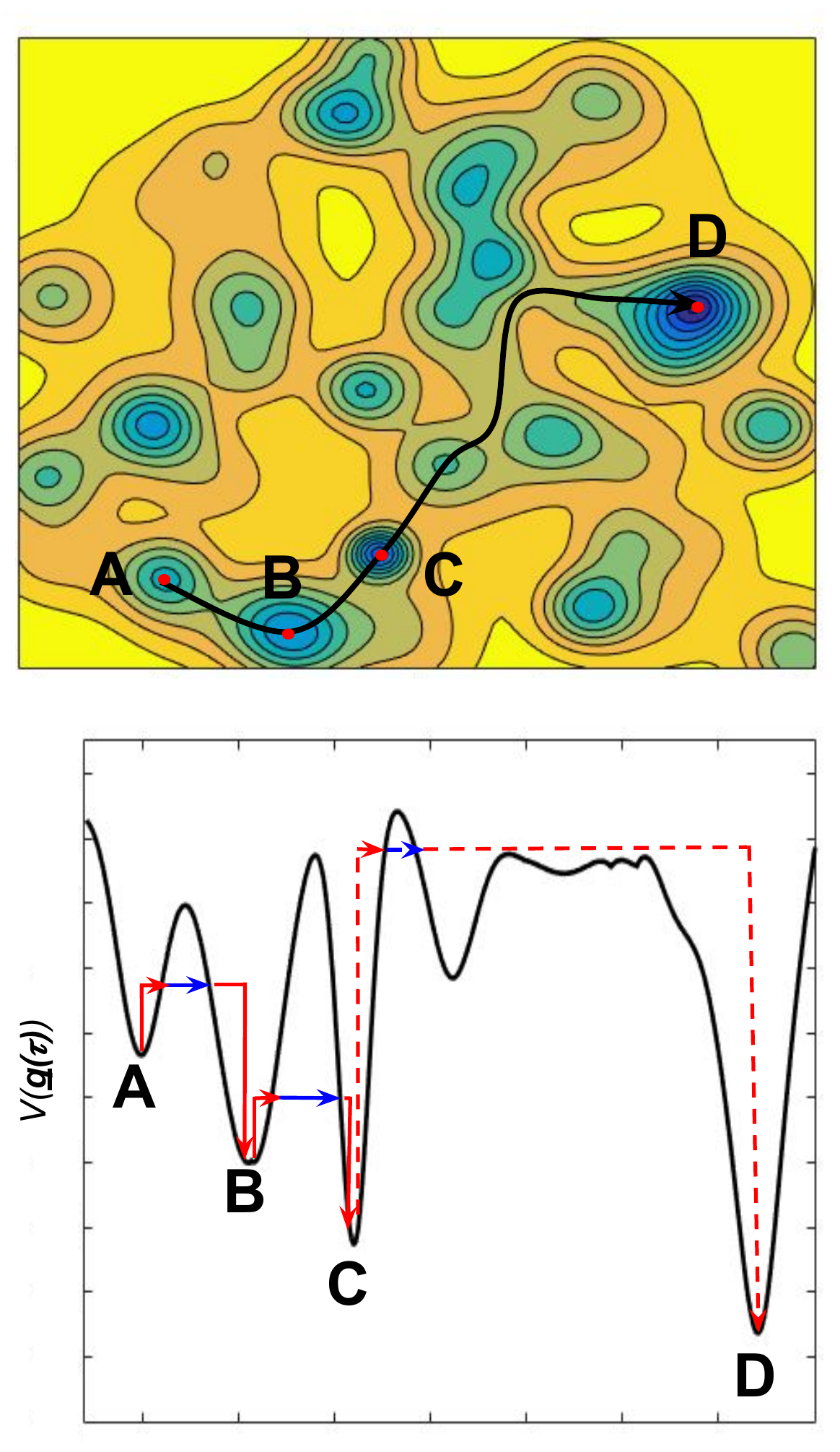}}
  \caption{{\bf Upper:} quantum annealer dynamics are
    dominated by paths through the mean field energy landscape that
    have the highest transition probabilities.  Shown is a path that
    connects local minimum A to local minimum D. {\bf Lower:} the
    mean field energy $V(q(\tau))$ along the path from A to D, as
    defined by Eqs.~\eqref{eq:V} and~\eqref{eq:q}. Finite range, thermally assisted tunneling can be thought of as a transition consisting of three steps: I. The system picks up thermal energy from the bath (red arrow up); II. The system performs a tunneling transition between the entry and exit points (blue arrow); III. The system relaxes to a local minimum by dissipating energy back into the thermal bath (red arrow down). 
    In transitions A $\to$ B  or B $\to$ C finite range tunneling considerably reduces the thermal activation energy needed to overcome the barrier.
    For long distance transitions in the lower part of the energy spectrum, such as C $\to$ D, the transition rate is still dominated by the thermal activation energy and the increase in transition rate brought about by tunneling is negligible. 
  }
\label{Thermal_Assisted_Tunneling}
\end{figure}

Because tunneling is more advantageous when energy barriers are tall and narrow, we expect this resource to be most valuable in the upper part of the energy spectrum. For instance, a random initial state is likely to have an energy well above the ground state energy for a difficult optimization problem such as the one in Eq.~\eqref{eq:problem}. However, the closest lower energy local minimum will often be less than a dozen spin flips away. Nevertheless, the energy barriers separating these minima may still be high. In such situations, if the transverse field is turned on to facilitate tunneling transitions, the transition rate to lower energy minima will often increase. By contrast, once the state reaches the low part of the energy spectrum, the closest lower local minimum is asymptotically $N$ spin flips away~\cite{santoro_theory_2002,Altshuler13072010,Farhi2012,knysh2015}. There, finite range tunneling may assist by effectively ``chopping off'' narrow energy ridges near the barrier top but the transition probability is still largely given by the Boltzmann factor. This description illustrates that finite range tunneling can be useful to quickly reach an approximate optimization, but will not necessarily significantly outperform SA when the task is to find the ground state (see Fig.~\ref{Thermal_Assisted_Tunneling}). 

The canonical QA protocol initializes the system in the symmetric superposition state, $\ket{+}^{\otimes N}$, which is the ground state at $t=0$. By a similar argument, we expect that finite range tunneling will drive the system adiabatically across energy gaps associated with narrow barriers,  preventing transitions to higher energies. However, in general, finite range tunneling will not be able to prevent Landau-Zener diabatic transitions for very small gaps resulting from emerging minima in the energy landscape separated by a wide barrier. This will often include the gap separating the ground state from the first excited state~\cite{santoro_theory_2002,Altshuler13072010,Farhi2012,knysh2015}.

The paper is organized as follows: In Section~\ref{wsc_results} we present our main results consisting of benchmarking the D-Wave 2X processor against SA and QMC on a crafted problem with a rugged energy landscape; Section~\ref{wsc_theory} introduces the theory of instantons in multi-spin systems, discusses tunneling simulation in QMC, and presents numerical results comparing QA and QMC for the ``weak-strong cluster pair" problem; Section~\ref{npp_kspin} presents numerical studies of generic problems with rugged energy landscapes that can potentially benefit from QA; Section~\ref{kspin_challenges} discusses the challenges associated with designing future annealers of practical relevance; and Section~\ref{summary} concludes with an overview and discussion. Further technical details can be found in Appendix~\ref{app:modelling} and~\ref{app_b}.

\section{Benchmarking physical quantum annealing on a crafted problem with a rugged energy landscape}\label{wsc_results}

In this section, we consider a problem designed so that finite cotunneling transitions of multiple spins strongly impacts the success probability. The previous section outlined several reasons why QA has a chance to outperform SA for problems with a rugged energy landscape.

\begin{figure}[h]
  \includegraphics[width= \columnwidth]{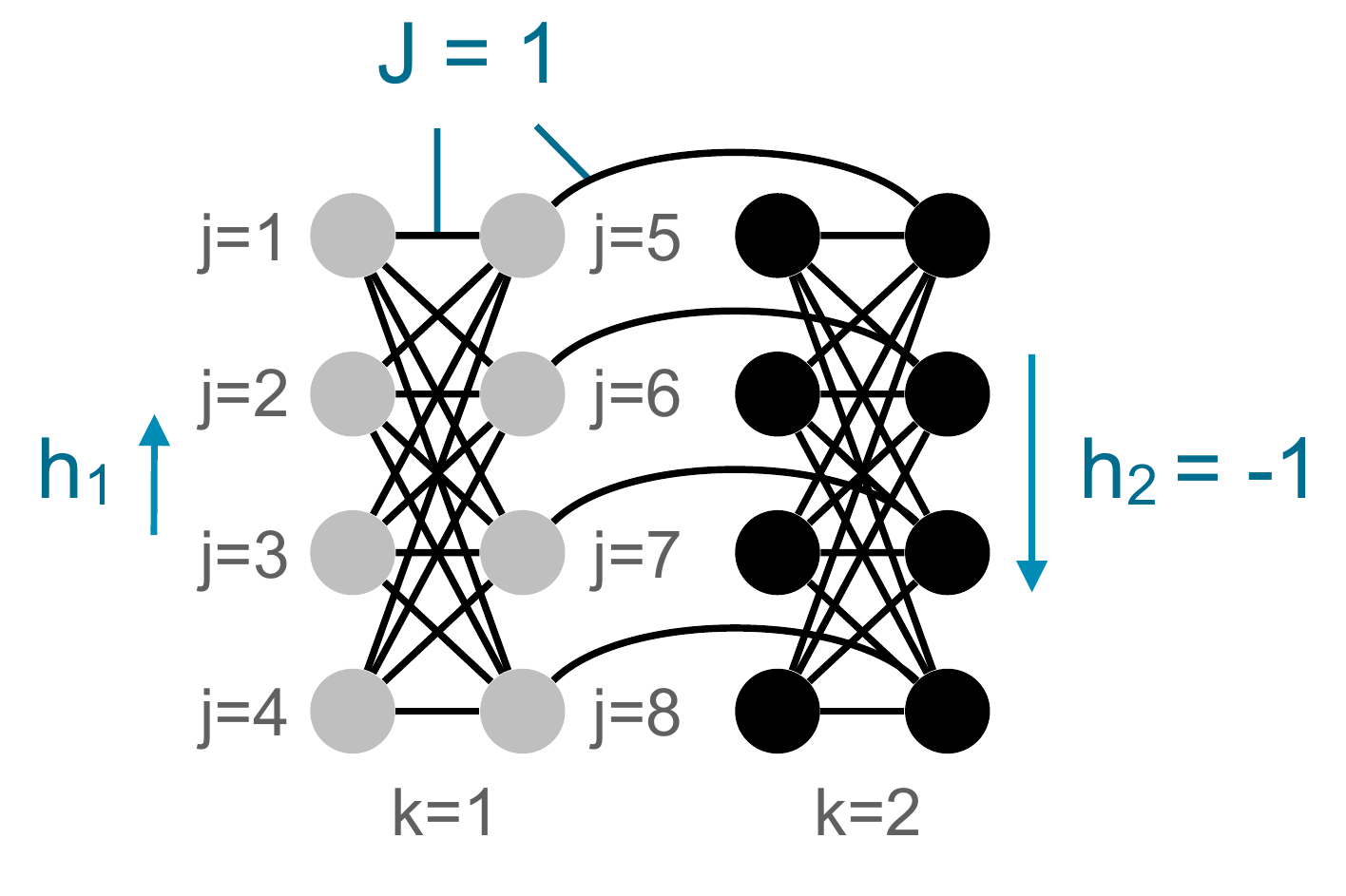}
  \caption{A pair of weak-strong clusters, consisting of 16 qubits in
    two unit cells of the Chimera graph. All qubits are
    ferromagnetically coupled and evolve as part of two distinct qubit
    clusters. At the end of the annealing evolution, the right cluster
    is strongly pinned downwards due to strong local fields acting on
    all qubits in that cell. However, the local magnetic field $h_1$ in the
    left cluster is weaker and serves as a bifurcation
    parameter. For $h_1 =0.44 < 1/2$ the left cluster will reverse its
    orientation during the annealing sweep and eventually align itself
    with the right cluster.}
\label{fig:Weak_Strong}
\end{figure}

In previous work we proposed a problem consisting of a pair of strongly connected spins (called clusters) to study the presence of functional cotunneling in D-Wave processors~\cite{boixo_computational_2014,Boixo2016}. Each cluster coincides with a unit cell of the native hardware graph, known as the Chimera graph. The problem Hamiltonian $H_{\rm P}$ in Eq.~\eqref{eq:H}  is of Ising form
\begin{align}
H_{\rm P}&=H_{\rm P}^{1} + H_{\rm P}^{2} + H_{\rm P}^{1,2}\label{eq:Hp_total}\\
H_{\rm P}^{k}&= -J \sum_{\langle j, j' \rangle \in \rm intra}\sigma_{k,j}^{z}\sigma_{k,j\prime}^{z}-\sum_{j=1}^{8}h_{k}\,\sigma_{k,j}^{z} 
\label{eq:Hp1}\\
H_{\rm P}^{1,2}&= -J \sum_{j \in \rm inter} \sigma_{1,j}^{z} \sigma_{2,j}^{z} \;.   \label{eq:Hp2a}
\end{align} 
All the couplings are ferromagnetic with $J=1$. The index $k \in \{1,2\}$ indexes a unit cell of the Chimera graph, the first sum in (\ref{eq:Hp1})  goes over the \emph{intra}-cell couplings
depicted in Fig.~\ref{fig:Weak_Strong}, the second sum  goes over the \emph{inter}-cell couplings corresponding to $j=5,6,7,8$ in Fig.~\ref{fig:Weak_Strong}, and $h_k$ denotes the local fields within each cell.

The local fields $0<h_1< 0.44$ (weak cluster) and $h_2=-1$ (strong cluster) are equal for all the spins within the cell. In this parameter regime, all spins of both clusters will point along the direction of the strong local fields in the ground state of the problem Hamiltonian $H_{\rm P}$. However, in the initial phase of the annealing evolution, all spins in the weak cluster orient themselves by following the local field in the opposite direction. At a later stage of the annealing evolution, the pairwise coupling between clusters becomes dominant; however, in the mean field picture, the state of the weak cluster is driven into a local minimum. Using a noise model with experimentally measured parameters for the D-Wave 2X processor, we numerically verify that the most likely mechanism by which all spins arrive at the energetically more favorable configuration is multi-qubit cotunneling (see Ref.~\cite{boixo_computational_2014} and Appendix~\ref{app:modelling}).

\begin{figure*}
  \centering
  {\includegraphics[width=\textwidth]{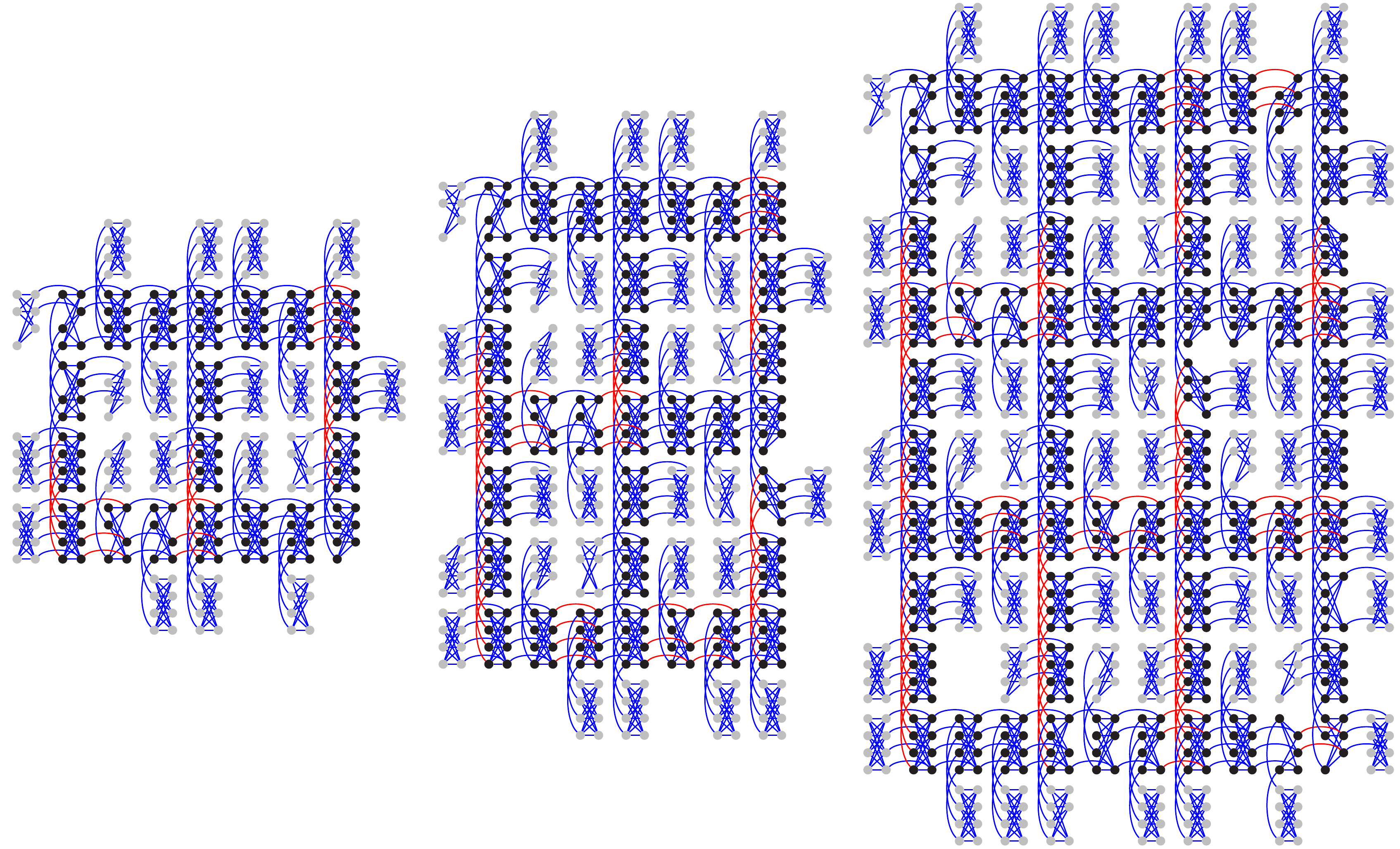}}
  \caption{Layout of several instances of the weak-strong cluster network problem on the D-Wave 2X processor. Shown are three different sizes with 296, 489 and 945 qubits. Each cluster consists of an eight qubit unit cell of the Chimera graph. Black dots depict qubits subject to a strong local field $h_R=-1$ while the gray dots represent qubits with the weak field $h_L=0.44$. Blue lines correspond to strong ferromagnetic couplings ($J=1$) and red lines to strong antiferromagnetic couplings ($J=-1$). Note that the graphs are somewhat irregular due to the fact that not all 1152 qubits are operational.}
\label{Glass_Layouts}
\end{figure*}

Using the weak-strong cluster pairs as building blocks, larger problems are formed by connecting the strong clusters to one another in a glassy fashion. That is, the four connections between two neighboring strong clusters are all set either to $+1$ (ferromagnetic) or $-1$ (antiferromagnetic), at random. With this procedure, we define a large class of instances having any size that we refer to as the ``weak-strong cluster networks'' problem. Fig.~\ref{Glass_Layouts} shows several examples of the layout of instances that were used in these benchmark tests~\footnote{The complete set of instances is available at\\ \url{http://goo.gl/yYKcb1}}. Due to the fact that not all of the qubits in the D-Wave 2X processor are properly calibrated, and hence available for computation, the instances become somewhat irregular. 

\subsection{D-Wave versus Simulated Annealing}

\begin{figure}[t]
  \centering
  {\includegraphics[width=\columnwidth]{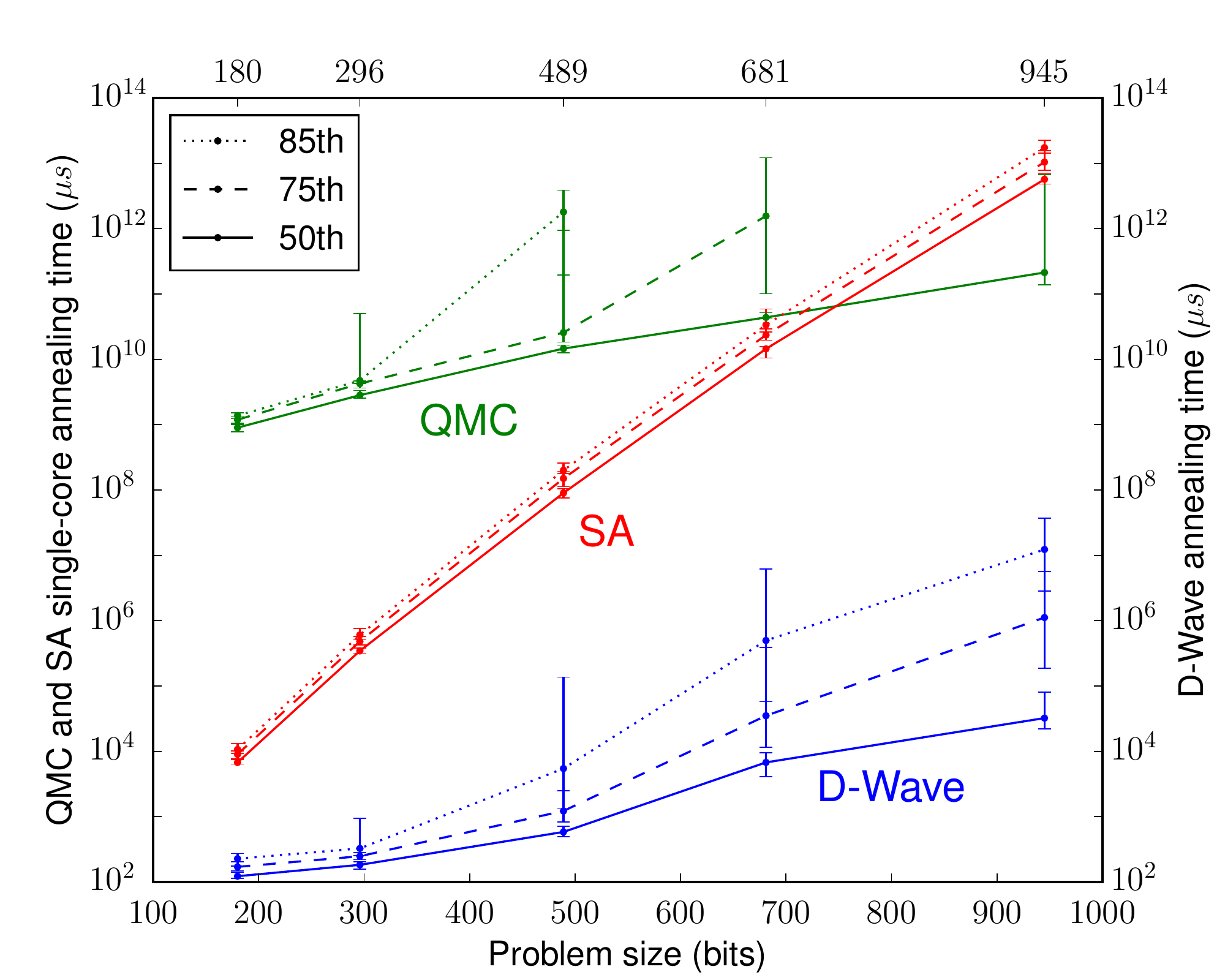}}
  \caption{Time to find the optimal solution with 99\% probability for different problem sizes. We compare Simulated Annealing (SA), Quantum Monte Carlo (QMC) and the D-Wave 2X. To assign a runtime for the classical algorithms we take the number of spin updates (for SA) or worldline updates (for QMC) that are required to reach a 99\% success probability and multiply that with the time to perform one update on a single state-of-the-art core. Shown are the 50th, 75th and 85th percentiles over a set of 100 instances. The error bars represent $95\%$ confidence intervals from bootstrapping. This experiment occupied millions of processor cores for several days to tune and run the classical algorithms for these benchmarks. The runtimes for the higher quantiles for the larger problem sizes for QMC were not computed because the computational cost was too high. For a similar comparison with QMC with different parameters please see Fig.~\ref{QMC_linear_versus-D-Wave_Timings}.}
\label{SA_QMC_versus-D-Wave_Timings}
\end{figure}

We now compare the total annealing time of SA to the total annealing time of the D-Wave 2X processor. Fig.~\ref{SA_QMC_versus-D-Wave_Timings} shows the time to reach the ground state with 99\% success probability, as a function of problem size for different quantiles. 

For D-Wave, we fix the annealing time at $20$ $\mu$s, the shortest time available due to engineering compromises, and estimate the probability $p_j$ of finding the ground state for a given instance $j$~\footnote{We take the mean over 16 gauges.}. Shorter times are optimal in this benchmark, as explained in Appendix ~\ref{app:modelling} and Ref.~\cite{boixo_computational_2014}. The total annealing time to achieve $p_j = 0.99$ is $20 \,\mu s \times \log(1-0.99)/\log(1-p_j)$. See Appendix~\ref{app:modelling} for details of the physical QA parameters (see also~\cite{harris_experimental_2010}).

We measure the computational effort of SA in units of runtime on a single core, which is natural when using server centers and convenient in order to compare to other classical approaches. Of course, the total runtime can be shortened by parallelizing the overall computation. Part of that process is embarrassingly parallel since SA finds its best solutions not in a single run, but by restarting from random bit configurations. Every restart could be executed on a different core. In fact, we used this strategy for our numerical benchmark. We find that median-case instances required $10^9$ independent runs (with $945 \times 5\cdot 10^4$ spin updates each) to find the optimal solution with 945 variables, on average.

We note that for instances with more qubits, more quantum hardware resources are brought to bear and therefore a fair comparison needs to take this into account~\cite{boixo_evidence_2014,ronnow_defining_2014}. As an extreme example, one could contemplate building special purpose classical hardware that would update as many spins in parallel as possible at state-of-the-art clock rates. The sets of spins that could be updated in parallel depends on the connectivity graph. Though such considerations are reasonable, we do not explore this possibility here as we believe that future quantum annealers will have higher connectivity which will severely limit the usefulness of such parallelism.

When estimating runtimes for SA, we follow the protocol laid out by Refs.~\cite{isakov_optimised_2015,boixo_evidence_2014,ronnow_defining_2014} and tuned SA for every problem size and quantile. Tuning means that the starting and end temperature, as well as the number of spin updates and the number of restarts are optimized to achieve a short overall runtime. We first measure the computational effort in units of sweeps (one sweep attempts to update all the spins). These times are plotted is $n_{\rm sweeps} \times N \times T_{\rm su}$, where $N$ is the number of spins. We used a spin update time $T_{\rm su} = 1/5$ ns (see Ref.~\cite{boixo_evidence_2014}).

Our key finding in this comparison is that SA performs very poorly on the weak-strong cluster networks. The D-Wave 2X processor is $1.8 \cdot 10^8$ faster at the largest size instances we investigated, which consisted of 945 variables. This problem was specifically engineered to cause the failure of SA: as explained above, the ``weak-strong cluster networks'' problem is intended to showcase the performance of annealers on a problem that benefits from finite range cotunneling. By contrast, the random Ising instances studied in Refs.~\cite{boixo_evidence_2014,ronnow_defining_2014} have only low energy barriers, as explained in Ref.~\cite{katzgraber_glassy_2014}.

\subsection{D-Wave versus Quantum Monte Carlo}\label{sec:d-wave_vs_qmc}

Next we compared the performance of path integral Quantum Monte Carlo
(QMC) with that of D-Wave for the same benchmark. QMC samples the
Boltzmann distribution of a classical Hamiltonian which approximates
the transverse field Ising model. In the case of a 2-spin model, the
discrete imaginary time QMC classical Hamiltonian is
\begin{multline}
    H_{\rm cl} = -\sum_{\tau=1}^M \(\sum_{jk} \frac{J_{jk}} M \sigma_j(\tau) \sigma_k(\tau) \right.\\
    \left. +   J^\perp(s) \sum_j \sigma_j(\tau) \sigma_j(\tau+1)\)\;,\label{eq:h_qmc}
\end{multline}
where $\sigma_j(\tau) = \pm 1$ are classical spins, $j$ and $k$ are site indices,
$\tau$ is a replica index, and $M$ is the number of replicas. The coupling between replicas is given
by
\begin{align}
  J^\perp(s) = - \frac {1} {2 \beta} \ln \tanh \frac {A(s) \beta} M\;,
\end{align}
where $\beta$ is the inverse temperature. The configurations for a
given spin $j$ across all replicas $\tau$ is called the worldline of
spin $j$. Periodic boundary conditions are imposed between $\sigma_j(M)$ and $\sigma_j(1)$. We used continuous
path integral QMC, which corresponds to the limit $\Delta \tau \to
0$~\cite{rieger_application_1998}, and, unlike discrete path integral
QMC, does not suffer from discretization errors of order $1/M$.

We numerically compute the number of sweeps $n_{\rm sweeps}$ required for QMC
to find the ground state with 99\% probability at different
quantiles. In our case, a sweep corresponds to two update attempts for each worldline. The computational
effort is $n_{\rm sweeps} \times N \times T_{\rm worldline}$, where
$N$ is the number of qubits and $T_{\rm worldline}$ is the time to
update a worldline. We average $T_{\rm worldline}$ over all the steps
in the quantum annealing schedule; however the value of $T_{\rm worldline}$ depends on the particular schedule chosen.
 As explained above for SA, we report the total computational effort of QMC in standard units of time
per single core. For the annealing schedule used in the current D-Wave
2X processor, we find
\begin{equation}
T_{\rm worldline}= \beta \times 870 \;{\rm ns} \label{eq:Twl_linear}
\end{equation} 
using an Intel(R) Xeon(R) CPU E5-1650 @ 3.20GHz. 

This study is designed to explore the utility of QMC as a classical optimization
routine. Accordingly, we optimize QMC by running at a low temperature, $4.8$ mK. We also observe that QMC with open boundary
conditions (OBC) performs better than standard QMC with periodic boundary
conditions in this case~\cite{isakov2015understanding}; therefore, OBC
is used in this comparison. We further optimize the number of
sweeps per run which, for a given quantile, results in the lowest
total computational effort. We find
that the optimal number of sweeps for the 50th percentile at the largest
problem size is $10^6$. This enhances the ability of QMC to
simulate quantum tunneling, and gives a very high probability of
success per run in the median case, $p_{\rm success} = 0.16$.

All the qubits in a cluster have approximately the same orientation
in each local minimum of the effective mean field potential. Neighboring local minima typically correspond to
different orientations of a single cluster.  Here, tunneling time is dominated by a single
purely imaginary instanton and is described by Eq.~\eqref{eq:mean_wkb} below. It was recently demonstrated that, in this situation, the
exponent $a_{\rm min}/\hbar$ for physical tunneling is identical to
that of QMC~\cite{isakov2015understanding}. As seen in Fig.~\ref{SA_QMC_versus-D-Wave_Timings}, we do not find a substantial difference in the
scaling of QMC and D-Wave (QA). However, we find a very substantial computational overhead
associated with the prefactor $B$ in the expression
$T=B e^{D a_{\rm min}/\hbar}$ for the runtime. In other
words, $B_{\rm QMC}$ can exceed $B_{\rm QA}$ by many orders of
magnitude.  The role of the prefactor becomes essential in situations
where the number of cotunneling qubits $D$ is finite, i.e., is
independent of the problem size $N$ (or depends on $N$ very
weakly). Between some quantiles and system sizes we observe a prefactor advantage as high as $10^8$.

\subsection{D-Wave versus other Classical Solvers}

Based on the results presented here, one cannot claim a quantum speedup for D-Wave 2X, as this would require that the quantum processor in question outperforms the best known classical algorithm. This is not the case for the weak-strong cluster networks. This is because a variety of heuristic classical algorithms can solve most instances of Chimera structured problems much faster than SA, QMC, and the D-Wave 2X~\cite{hamze_fields_2012,selby_efficient_2014,zintchenko2015local} (for a possible exception see~\cite{hen2015probing,king_benchmarking_2015})~\footnote{Nevertheless, it is interesting to note that Ref.~\cite{ronnow_defining_2014} defines a \emph{limited quantum speedup} as ``a speedup obtained when comparing specifically with classical algorithms that correspond to the quantum algorithm in the sense that they implement the same algorithmic approach, but on classical hardware ... a natural example is quantum annealing implemented on a candidate physical quantum information processor versus either classical simulated annealing, classical spin dynamics, or simulated quantum annealing.'' In the weak-strong cluster networks benchmark we show that D-Wave 2X outperforms these three algorithms. We find a substantial advantage against QMC in the prefactor, not the scaling. For the comparison with classical spin dynamics please see Ref.~\cite{boixo_computational_2014}}. For instance, the Hamze-de Freitas-Selby algorithm~\cite{hamze_fields_2012,selby_efficient_2014}, performs a greedy sequence of random large neighborhood optimizations. Each large neighborhood is defined by first replacing each 4-qubit column in a cluster with a large spin, and then expanding a tree of large spins which covers $\sim$ 77\% of all spins (and more than half of all $8$ qubit clusters). In the particular case of a weak-strong cluster pair this algorithm avoids the formation of the energy barrier.

However, we believe that such solvers will become ineffective for the next generation of annealers currently being designed. The primary motivation for this optimism is that the connectivity graphs of future annealers will have higher degree. For example, we believe that a 10-dimensional lattice is an engineerable architecture. For such dense graphs, we expect that cluster finding will become too costly. On the contrary, multi-qubit cotunneling is a general phenomenon in spin glasses that is not limited to sparse graphs. 

We have also learned from the Janus team, working with special purpose
FPGAs to thermalize Ising models on a $32^3$ cube, that they found
cluster finding not to be helpful~\footnote{Victor Martin-Mayor,
  personal communication.}.

\subsubsection*{A Remark on Scaling}

Certain quantum algorithms have asymptotically better scaling than the best known classical algorithms. While such asymptotic behavior can be of tremendous value, this is not always the case. Moreover, there can be substantial value in quantum algorithms which do not show asymptotically better scaling than classical approaches. The first reason for this is that current quantum hardware is restricted to rather modest problem sizes of less than order 1000 qubits. At the same time, when performing numerical simulations of quantum dynamics, in particular when doing open system calculations, we are often limited to problem sizes smaller than 100 qubits. Extrapolating from such finite size findings can be misleading as it is often difficult to determine whether a computational approach has reached its asymptotic behavior. 

When forecasting the future promise of a given hardware design, there is a tendency to focus on the qubit dimension. However, this perspective is not necessarily helpful. For instance, when looking at runtime as a function of qubit dimension, one may conclude that the measurements reported here indicate that QMC and physical annealing have a comparable slope, scale similarly and that therefore, the upside of physical annealing is bounded.  However, the large and practically important prefactor depends on a number of factors such as temperature. Furthermore, we expect future hardware to have substantially better control precision, substantially richer connectivity graphs and dramatically improved $T_1$ and $T_2$ times. With such changes, next generation annealers may drastically increase the constant separation between algorithms, leading to very different performance from generation to generation. To illustrate how dramatic this effect can be, when we ran smaller instances of the weak-strong cluster networks on the older D-Wave Vesuvius chips we predicted that at 1000 variables D-Wave would be $10^4$ times faster than SA. In fact, we observed a speedup of more than a factor of $10^8$. This is because certain noise parameters were improved and the new dilution refrigerator operates at a lower temperature. Similarly, we suspect that a number of previous attempts to extrapolate the D-Wave runtimes for 1000 qubits will turn out to be of limited use in forecasting the performance of future devices. For this reason, the current study focuses on runtime ratios that were actually measured on the largest instances solvable using the current device, rather than on extrapolations of asymptotic behavior which may not be relevant once we have devices which can attempt larger problems.

\section{Spin Cotunneling in QA and QMC}\label{wsc_theory}

\subsection{Instantons in systems with multiple spins}
Cotunneling consists of system state transitions in which a group of spins simultaneously change their orientation with energy well below the energy of the (mean field) potential barriers. Tunneling is a quintessential quantum phenomenon; real time dynamics of classical trajectories cannot describe barrier penetration when the system wavefunction extends to classically forbidden regions. In such situations, the exponential decay of the wavefunction under the barrier is often captured through the path integral formalism by computing the minimum action of the trajectories in imaginary time~\cite{coleman_fate_1977,0038-5670-25-4-R01}. This approach was also extended to treat the tunneling of large magnetic moments with conserved total spin~\cite{Chudnovsky-book}.

Tunneling in  mean field spin models  can be described using the path integral over spin-coherent states in imaginary time~\cite{nagaosa2013quantum}.  The tunneling path connects the minima of the mean field  potential 
\begin{equation}
V(\um ,t)=\langle \Psi_{\um} | H(t)| \Psi_{\underline{\bf m}}\rangle\;,\label{eq:V}
\end{equation}
where $H(t)$ is the time-dependent QA Hamiltonian from Eq.~\eqref{eq:H} and $| \Psi_{\underline{\bf m}}\rangle$ is a product state
 \begin{align}
| \Psi_{\underline{\bf m}}\rangle =  \bigotimes_j \left[  \cos \frac {\theta_j} 2 \ket 0 + e^{-i \phi_j} \sin \frac {\theta_j} 2 \ket 1\right ]\; \;.\label{eq:m}
 \end{align}
The coherent state of  the $j$-th spin is defined by a vector on the Bloch  sphere 
 \begin{equation}
 {\bf n}_j =(\sin\theta_j \cos\phi_j,  \sin\theta_j \sin\phi_j, \cos\theta_j)\;.
\end{equation}
and the corresponding state of the $N$-spin system is defined by the vector 
$\um=({\bf n}_1,{\bf n}_2,\ldots,{\bf n}_N)$.

Towards the beginning of a QA evolution, the system remains near $\um_0(t)$, the global minimum  of the time-dependent potential $V(\um,t)$, which connects to the global minimum at the initial time. 
Later on, $V(\um,t)$ undergoes a bifurcation, which may cause the initial minimum to become metastable. At that point, the system may be able to tunnel to the new global minimum $\um_1(t)$ when $V(\um_0)\simeq V(\um_1)$.
Here we omit the argument $t$, whose value corresponds approximately to the moment when the minima exchange order. Such tunneling events are sometimes accompanied by thermal activation if QA is performed at finite temperatures (i.e. thermally assisted tunneling)~\cite{KostyaVadim2015}. The sequence of bifurcations and associated tunneling events can continue multiple times before the global minimum of $H_{\rm P}$ is reached.

Mathematically, the tunneling process can be described as an evolution of a spin system
along the instanton path 
\begin{equation}
\uq(\tau)=({\bf n}_1(\tau),{\bf n}_2(\tau),\ldots,{\bf n}_N(\tau))\label{eq:q}
\end{equation} over imaginary time $\tau\in(0,\beta)$ where $\beta=\hbar/k_B T$ and  $T$ is the system temperature.
The details of this analysis will be provided elsewhere \cite{ID:2016}. Here, we simply outline the main argument.
 The  path component for  the $j$-th spin is described by a vector ${\bf n}_j(\tau)$
 with periodic boundary conditions ${\bf n}_j(0)={\bf n}_j(\beta)$.  The vector  ${\bf n}(\tau)$ is complex  (see Eqs.~(67),(68) in the Supplementary Material of Ref.~ \cite{isakov2015understanding}) and can be written in the form 
  \begin{equation}
 {\bf n} =(\sin\theta_j \cosh\varphi_j,  -i \sin\theta_j \sinh\varphi_j, \cos\theta_j)\;,\label{eq:n}
\end{equation}
corresponding to a purely imaginary azimuthal angle $\phi_j(\tau)=-i \varphi_j(\tau)$. The initial point of the  instanton $\uq(0)$ corresponds to the initial minimum  of $V(\um )$. The   midpoint of the trajectory $\uq(\beta/2)$  typically corresponds to the exit point of the potential barrier in the vicinity of the final minima
 \begin{equation}
 \uq(0)= \uq(\beta)\simeq \um_0,\quad \uq(\beta/2) \simeq  \um_1.\label{eq:tp}
 \end{equation}
 
We first consider the  simplified situation where the domain of $D$ spins tunnel simultaneously. The total spin of the tunneling domain is conserved and all spins in the domain move identically through the instanton trajectory,  ${\bf n}_j(\tau) \equiv {\bf n}(\tau)$ for all $j\in[1,D]$. Then, the mean field potential for the instanton can be rescaled as
 \begin{equation}
 V(\uq(\tau))=D \upsilon( {\bf n}(\tau))\;.\label{eq:Vup}.
 \end{equation}
 In the limit of low temperatures $\beta\rightarrow \infty$, the instanton action takes the form $ S[{\bf n}(\tau)]=D a[{\bf n}(\tau)]$ with
\begin{equation}
 a[{\bf n}(\tau)]=\frac{\hbar }{2}\,\omega[{\bf n}(\tau)]+ \int_{0}^{\infty} d\tau  \upsilon[{\bf n}(\tau)]\;,\label{eq:a}
 \end{equation} 
 where $\omega$ describes a ``Berry phase'' type  contribution 
 \begin{equation}
 \omega[{\bf n}(\tau)]=\int_{0}^{\infty}d\tau (1-\cos\theta(\tau))\dot \varphi(\tau)\;.\label{eq:Berry}
 \end{equation}
 The exponential dependence of the tunneling rate,
 \begin{equation}
 W_{\rm QA}\sim e^{-D a_{\rm min}/\hbar},\quad
 a_{\rm min}=\min_{\bf n(\tau)} a[{\bf n}(\tau)]\;,
  \label{eq:W}
 \end{equation}
is given by the minimum of the rescaled action.
  
We now consider the case when the tunneling of the spin domain is enabled by the transverse field and the Hamiltonian is of the type in Eq.~(\ref{eq:H}),
\begin{equation}
H=-A \sum_{j=1}^{N}\sigma_{j}^{x}- B H_{P}^{\rm cl}\left (\sigma_1^z,\ldots,\sigma_D^z \right)\;,\label{eq:HD1}
\end{equation}
where $H_{P}^{\rm cl}\left (s_1,\ldots,s_D \right)$ is a classical cost function of  binary variables $s_k=\pm 1$. Assuming that
$H_{P}^{\rm cl}=H_{P}^{\rm cl}(\sum_{j=1}^{D}\sigma_{j}^{z})$, and the system is in a state of maximum total spin, all spins will tunnel together. Thus,
\begin{equation}
\upsilon( {\bf n}(\tau))=-A \sin\theta(\tau) \cosh\varphi(\tau)-B g(\cos\theta(\tau)),
\end{equation}
  where the function $g(x)=D^{-1} H_{P}^{\rm cl}(D x)$  is the rescaled mean field potential energy of the spin
system. At zero temperature, the minimum action can be written in a more familiar form (cf. Supplementary Material in Ref.~ \cite{isakov2015understanding} and also Ref.~\cite{Garg-wkb}),
 \begin{equation}
\frac{a_{\rm min}}{\hbar}=\int_{\theta_0}^{\theta_1} {\rm arcsinh }
\left(\frac{v(\theta)}{A\sin\theta} \right) \sin\theta
d\theta\label{eq:mean_wkb},
\end{equation}
where $v(\theta)= \sqrt{B^2(g(\cos\theta)-g(\cos\theta_0))^2-A^2\sin^2\theta}$ is a linear velocity along the instanton path and the angles $\theta_{i}$ correspond  to the minima of the potential $\upsilon( {\bf n})$ (the values of $\sinh \varphi_j=0$ at the minima).

In general, in mean field spin models, the total spin is not conserved  throughout the instanton trajectory (\ref{eq:q}). The corresponding action~\cite{nagaosa2013quantum} can be written in a form that is a direct generalization of the simplified case (\ref{eq:a})
 \begin{equation}
 S[\uq(\tau)]=\frac{1}{2}\sum_{i=1}^{D} \omega[{\bf n}_j(\tau)]+\int_{0}^{\beta} d\tau\,  V[\uq (\tau)]\;,\label{eq:S1}
 \end{equation} 
 where  $\omega$ is given in Eq.~(\ref{eq:Berry}). If the azimuthal angles are purely imaginary, as discussed above, the action is real. The dominant contribution to the path integral will be given by the instanton path with the least action
  \begin{equation}
  W_{\rm QA}\sim e^{-S_{\rm min}/\hbar},\quad
 S_{\rm min}=\underset{\uq(\tau)}{\mathrm{min}} S[\uq(\tau)]\;.
  \label{eq:W1}
 \end{equation}
 In general, the action in Eq.~(\ref{eq:W1}) also grows linearly with the
 number of cotunneling spins as in the simplified case of Eq.~(\ref{eq:W}) because the individual spin contributions to the action are highly correlated at the instanton trajectory. Accordingly, we may define $S_{\rm min} = D a_{\rm min}$.

\subsection{Tunneling simulation in QMC}

In the path integral QMC algorithm one introduces an extra dimension
associated with the imaginary time axis in order to simulate the
multispin tunneling phenomenon on a classical computer, as seen
above. This is done by using a Suzuki-Trotter decomposition and
representing the partition function of the system $Z$ in terms of the
path integral over the spin trajectories
$\underline{\sigma}(\tau)=\{\sigma_j(\tau)\}_{j=1}^{N}$, see
Eq.~\eqref{eq:h_qmc}. These trajectories are periodic along the
imaginary time axis $\sigma_j(0)=\sigma_j(\beta)$. For each spin $j$,
the set of values $\sigma_j(\tau)=\pm 1$ form a path component
referred to as a worldline. The time step along the worldline
is $\Delta\tau=\beta/M$ where $M$ is the number of Trotter slices
(i.e. the number of spin replicas in the worldline). Sampling the system states
along this extra dimension introduces an additional overhead in
classical computation that does not exist for the corresponding quantum dynamics.

The runtime $T_{\rm QMC}$ of QMC can be thought of as a product of three factors,
\begin{align}
T_{\rm QMC}&=N\,n_{\rm sweeps}\,T_{\rm worldline} \label{eq:TQMC}
\end{align}
where $N$ is the problem size that is equal to the number of worldlines. % The factor $n_{\rm sweeps}$  is the  number of  worldline updates during the algorithm execution divided by $N$.  
The number of sweeps $n_{\rm sweeps}$, in general, depends exponentially on the typical size $D$ of the cotunneling domain, $n_{\rm sweeps} \propto e^{\alpha D}$, where $\alpha = \alpha(\beta)$ also depends on the inverse temperature $\beta$. In cases where $D={\cal O}(N)$ the growth of this  factor  with $N$ reflects a major computational bottleneck of QMC and QA. As was shown recently by some of the authors~\cite{isakov2015understanding}, the exponent in QMC and QA is the same for a broad class of problems. 

According to the findings reported in this paper, the prefactor in $n_{\rm sweeps}$, along with the factors $N$ and $T_{\rm worldline}$ in $T_{\rm QMC}$, are significantly different for QMC and QA. The value of  $T_{\rm worldline}$ found in our simulations is given in Eq.~\eqref{eq:Twl}. We also find that 
\begin{equation}
n_{\rm sweeps} \gg {\rm 1 \, ns}/T_{\rm QA}\label{eq:cond1}\;,
\end{equation}
where $T_{\rm QA}$ is the duration of QA and we use a normalization factor of one nanosecond, corresponding to
the typical time scale of superconducting QA devices. We expect that the above relation will remain true even if the scaling of both quantities with $D a_{\min} /\hbar$ is the same.

\subsection{Comparison of QA and QMC for the ``weak-strong cluster pair'' problem}  

\begin{figure}[h]
  \centering
  \includegraphics[width=\columnwidth]{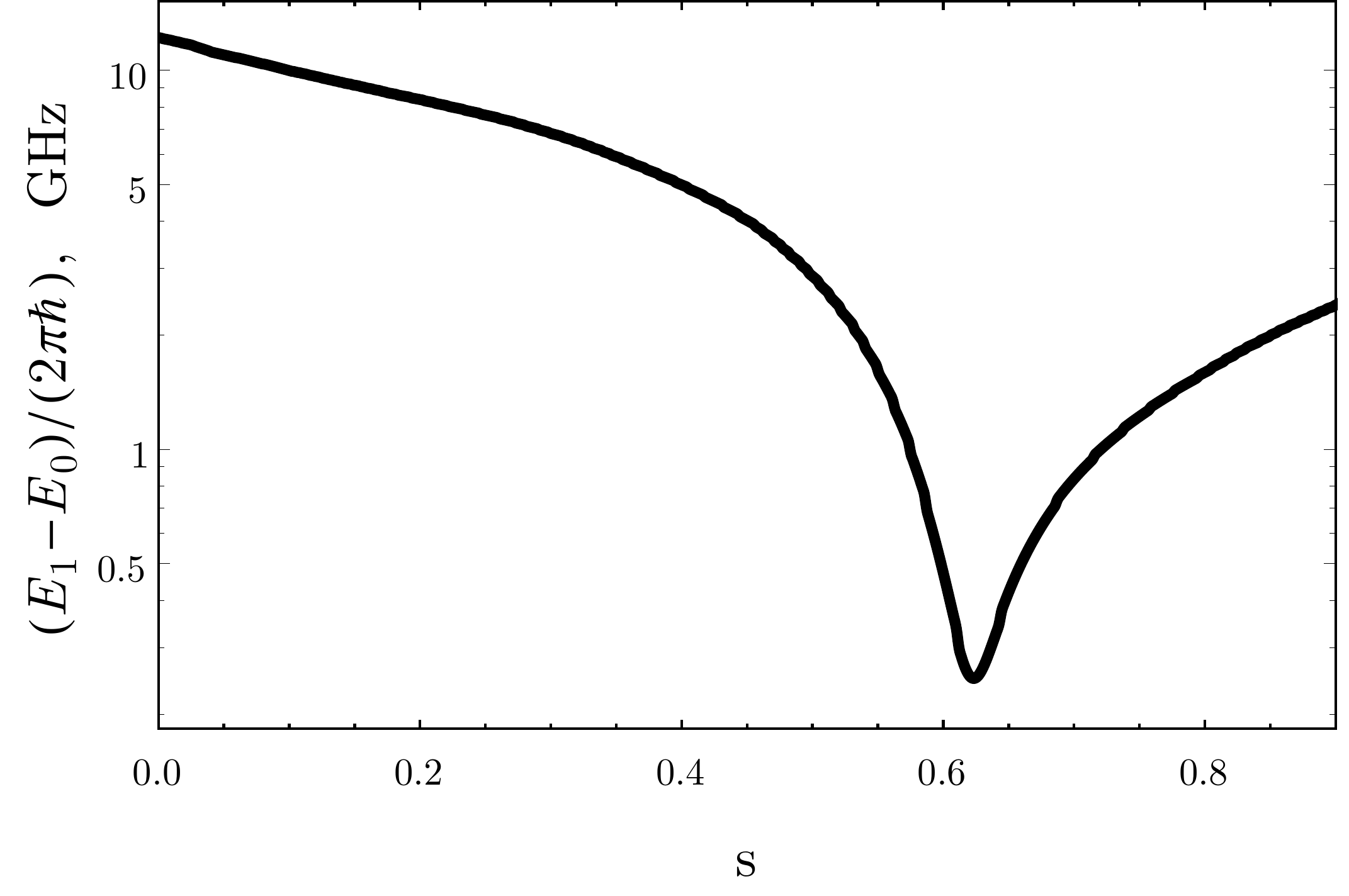}
  \caption{Gap of the quantum Hamiltonian for $h_1 = 0.44$ as a
    function of the annealing parameter. The solid  line is the
    energy difference between the ground state and the first excited
    state. The avoided crossing at $t/T_{\rm QA}=0.62$  corresponds to a minimum
    gap of 248 MHz.  The next excited state is separated by a gap in excess of 2 GHz.}
  \label{fig:gaps}
\end{figure}

We use the modeling considerations described above to compare QA and QMC in the system corresponding to the weak-strong cluster pair problem (see discussion in  Sec. II,  Fig.~\ref{fig:Weak_Strong}, and Eqs.~(\ref{eq:Hp_total}),(\ref{eq:Hp1}),(\ref{eq:Hp2a})).
Tunneling in this system corresponds to an avoided crossing between the two lowest energy levels of the Hamiltonian, shown in Fig.~\ref{fig:gaps}. All other levels lie high above the first two and are never excited. During tunneling, the total spin of the left cluster reverses orientation. The number of cotunneling spins is $D=8$ in this case while the total number of spins is $N=16$. 

Proper analysis of the tunneling probabilities and related QA success rates should also account for coupling to the environment. We studied the success probabilities in QA for the 2-cluster problem using the approach developed in  \cite{boixo_computational_2014}. The results are summarized in Fig.~\ref{fig:p0T} of   App.~\ref{app:modelling}.
Decreasing the temperature of QA compared to the temperature of the D-Wave device suppresses steeply the transition rates between the states because of the increase in the reconfiguration energy~\cite{amin_macroscopic_2008,harris_experimental_MRT_2008,boixo_computational_2014} (see Fig.~\ref{fig:W10T}). This, together with the suppression from the Boltzman factor in the transition rates, leads to an increase of the final success probability $p_0$ to find the ground state. Once the temperature reaches 5 mK, the success probability stays above 90\%, even at  QA schedules that are faster than $T_{\rm QA}$=300 nanoseconds.  On the other hand, adiabatic transitions near the avoided crossing are suppressed even at $T_{\rm QA}\simeq $ 71 nanoseconds, as can be seen from solutions to the time-dependent  Schr\"odinger  equation, shown in Fig.~\ref{fig:TDS}.

In the previous studies involving D-Wave devices (see \cite{boixo_computational_2014} for references) it was inferred from the data that the low frequency noise components of the spectral density, providing a leading contribution to the qubit  line-width $W$ (\ref{eq:par}),(\ref{eq:p0}), have effective frequency cutoffs much below 314 MHz (15 mK).   In the current QA schedule of 20 $\mu$s, the system spends only a small fraction of this time in the vicinity of the  avoided crossing where thermal excitations from the ground state are possible. For a QA schedule duration of $\sim$ 100 ns, we expect that the effective noise strength will be weaker than at the current schedule. This would lead to the suppression of  the thermal excitations from the ground state.

To compare  simulations of QA at a temperature of 5 mK with the QMC
performance, we choose a duration of QA such that the probability to
reach the ground state at the end of QA equals 0.95. As we discussed
above this can be achieved at
\begin{equation}
T_{\rm QA}=71 \,{\rm ns}\label{eq:TQAs}\;.
\end{equation}

In setting up the QMC simulations our objective is to select the two parameters, number of sweeps per qubit $n_{\rm sweeps}$ and $\beta$, to minimize $T_{\rm QMC}$  for a given probability of success  $p_0$ to find the system at the end of the QA in the ground  state where all spins point down. Essentially we need to minimize the product of $ \beta n_{\rm sweeps}$ keeping $p_0$ fixed.

\begin{figure}[h]
  \centering
  \includegraphics[width=\columnwidth]{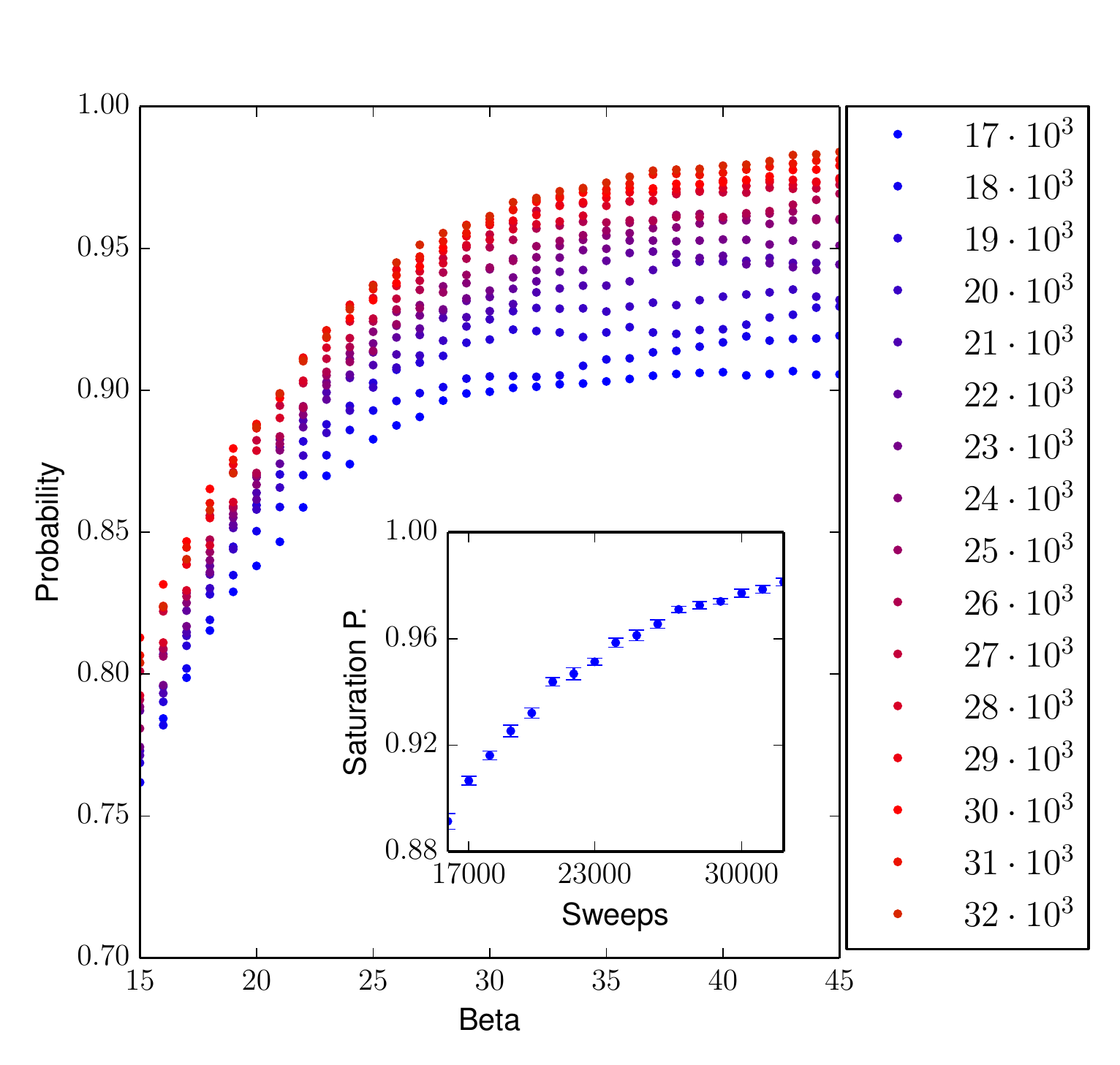}
  \caption{Probability of success versus $\beta$ for
    QMC with the D-Wave 2X schedule. Different colors correspond to
    different number of sweeps (see legend). The embedded plot shows
    the probability of success at saturation for different number of
    sweeps. We use periodic boundary conditions, which performed
    better than open boundary conditions in this case.}
  \label{fig:sweeps}
\end{figure}

In Fig.~\ref{fig:sweeps} we plot the success probability of QMC $p_0=p_0(\beta,n_{\rm sweeps})$ as a function of $\beta$ (inverse temperature) for different number of sweeps. We see that increasing $\beta$ increases $p_0$; however, the success probability saturates at some value
 \begin{equation}
p_0(\beta,n_{\rm sweeps}) \leq p^{\rm sat}_0(n_{\rm sweeps})\label{eq:p0max},
\end{equation}
which itself depends on the number of sweeps. The saturation probability $ p_0^{\rm sat}(n_{\rm sweeps})$ is plotted in the insert to  Fig.~\ref{fig:sweeps}. By fixing the success probability $p_0=0.95$ we select the optimal number of sweeps. Then, by looking on the main plot we determine the value of $\beta_{\rm sat}$ were saturation occurs.
 A detailed study shows that this procedure gives the minimum value of the product $ \beta n_{\rm sweeps}$. The optimal values are
 \begin{equation}
 n_{\rm sweeps}=23,000,\quad \beta=32.5\label{eq:opt}\;.
 \end{equation}
 The total time to update one worldline with the D-Wave 2X schedule is
 \begin{equation}
T_{\rm worldline}=  28.3 \,\mu{\rm s}  \label{eq:Twl1}
\end{equation} 
and the total runtime QMC per qubit, according to Eq.~(\ref{eq:TQMC}), is
\begin{equation}
\frac{T_{\rm QMC}}{N}=0.75 \,{\rm seconds}\;.\label{eq:TQMC2}
\end{equation}
By comparing this with Eq.~(\ref{eq:TQAs}), we estimate that
\begin{equation}
\frac{T_{QMC}/N}{T_{QA}}\sim  10^7  \quad (T_{\rm QA}=71\, {\rm ns},\quad T=5\, {\rm mK})  \label{eq:QE}.
\end{equation}
This ratio will need to be multiplied by the number of qubits to obtain the overall speed up factor (e.g., $\sim 10^{10}$ for 1000 qubits).

Implementing  fast QA schedules or operating flux qubits at 5 mK will require improvements in the  control electronics and other elements of the design. Furthermore, readout will need to be made much faster than in the current D-Wave devices.  However, the estimates presented above serve to emphasize the significant promise of QA, as compared to QMC when the system adiabatic evolution \lq\lq under the gap" becomes coherent and thermal excitations are suppressed.

\section{Numerical Studies of Quantum Annealing for Generic Problems with rugged Energy Landscapes}\label{npp_kspin}

Runtime advantages for the quantum processor described above are only valuable if they extend to problems of practical interest. While rather obvious, it may we worth delineating criteria for problems that are suitable for treatment with a quantum annealer:
\begin{enumerate}
\item Solutions to the problem are valuable or interesting.
\item The problem is representable on hardware that can be built in the near future.
\item Quantum annealing offers a runtime advantage.
\end{enumerate}

\subsection{Number Partitioning}

A valuable and interesting practical problem is the Number Partitioning Problem (NPP). The NPP is defined as follows: Given a set of $N$ positive numbers $(a_1, ..., a_N )$ find a partition ${\cal P}$ of this set into two groups that minimizes the partition residue $ E = | \sum_{j \in {\cal P}}  a_j- \sum_{j \notin {\cal P}} a_j | $.  A partition ${\cal P}$ can be encoded  by Ising spin  variables $s_j=\pm 1:\, s_j=+1$ if $j\in {\cal P}$ and $s_j=-1$ otherwise.  Thus, the NPP cost function is
\begin{equation}
E({\bf s})=|\Omega_{\bf s}|,\quad \Omega_{\bf s}=\sum_{j=1}^{N}a_j s_j\label{eq:NPP}
\end{equation}
where ${\bf s}=(s_1,\ldots,s_N)$ is a spin configuration and $\Omega_{\bf s}$ is a {\it  signed} partition  residue. Number partitioning is also one of Garey and Johnson's six basic NP-hard problems that lie at the heart of the theory of NP-completeness \cite{garey1979computers}. In studies of the average-case computational complexity of NPP, one usually assumes that $\{ a_1, ..., a_N \}$ are independent, uniformly distributed random numbers in the interval $[0,1)$.  The average-case complexity of NPP is exponential in $N$ when the number of bits $b$ used to represent the numbers $ a_j $ satisfies the condition: $b/N \geqslant \kappa_c \simeq 1-\frac{1}{2N}\log_2 N$ \cite{mertens1998phase}. Our focus is on hard instances of NPP and we will be studying the random NPP ensemble  with $b=N$. NPP has many practical applications including multiprocessor
scheduling  and the minimization of VLSI circuit size and delay, public key cryptography and others (see references in  \cite{mertens2001physicist}).

NPP is attractive for numerical studies because for $b/N>\kappa_c$, the typical runtime of all known algorithms for NPP scales exponentially with large coefficients in the exponent and often, the asymptotic behavior can already be seen at sizes as low as 20 variables. For our purposes, NPP is a useful problem to study in the context of quantum annealing because it possesses extremely rugged energy landscapes where a single bit flip can result in dramatic energy changes. Its low energy band resembles that of the random energy model as there is almost no correlation between the state and its energy \cite{mertens2000random}. The $2^N$ signed partition residues $\Omega$ can be thought of as drawn from the Gaussian distribution
\begin{equation}
p(\Omega)=\frac{2}{\sqrt{2\pi N  \langle a^2\rangle }}\exp\left(-\frac{\Omega^2}{2N  \langle a^2\rangle}\right); \label{eq:p}
\end{equation}
\noindent
where $  \langle a^2 \rangle = \frac{1}{N} \sum_{j=1}^{N} a_{j}^{2}$.   The distribution of the cost function values  $E=|\Omega |\geqslant$ is given by $2 p(E)$. 
By picking a bit string at random, one gets an  average  value of the cost function  $\langle E \rangle  = \sqrt{2 \langle a^2\rangle  N/\pi } $. The {\it minimum} value of this cost function is exponentially small, with median value  $E_{\rm min} \sim \langle E \rangle \,2^{-N}$ \cite{mertens2001physicist}.

An obvious heuristic algorithm for NPP starts by placing the largest number in one of the two subsets. The next largest number is then placed in the set whose elements sum to the smallest value and this continues until all numbers are assigned. The idea behind this greedy heuristic is to keep the discrepancy small with every decision. This gives the scaling of the resulting partition residue as ${\cal O}(1/N)$.
The differencing method of Karmarker and Karp \cite{mertens2001physicist}, also called the KK heuristics, is a polynomial time approximation algorithm. The key idea of this algorithm is to reduce the size of the numbers by replacing the two largest numbers by the absolute value of their difference. It has been proven \cite{yakir1996differencing} that the differencing method gives a minimum residue $E^{\rm KK}_{\rm min}$ such that 
\begin{equation}
E^{\rm KK}_{\rm min}\sim N^{-\alpha \log N},\quad \alpha=0.72\;.\label{eq:KK}
\end{equation} 

The time complexity of both greedy and KK heuristics is $N \log N$  \cite{mertens2001physicist}. The residual energies reached by both methods are much smaller than the average partition residue  $\langle E \rangle $, but far greater than the minimum residue $E_{\rm min} $. The absence of efficient heuristics for these hard cases is a particular feature of  NPP. It is attributed   to the extremely rugged energy landscape in the low part of the energy spectrum  \cite{mertens2001physicist}.  The statistics of the NPP energy landscape was studied analytically in \cite{smelyanskiy2002dynamics} and numerically in \cite{stadler2003phase}. 

This type of landscape leads to the exponential complexity of QA for NPP that was obtained in \cite{smelyanskiy2002dynamics} via direct solution of the time-dependent  Schr{\"o}dinger equation.
We observe that the particularly challenging instances of NPP violate the second condition of our ``suitability'' criteria as the numbers  $a_j \in (0, ..., 2^N-1) $ should be drawn from a set whose cardinality grows exponentially with $N$. This translates to a requirement that the bit precision for the coupling coefficients  $J_{jk}$ grow with $N^2$ if one were to express NPP as a quadratic binary optimization problem with objective function $\sum_{jk=1}^{N}a_j a_k s_j s_k$ corresponding to the form  (\ref{eq:problem}) with $K=2$. However, for numerical studies this is not a concern.

The runtime behavior of SA, QMC and QA, simulated using the Schr{\"o}dinger equation or other methods, is shown in Table~\ref{NPP_Table}. The relative performance of these algorithms applied to NPP is similar to their relative performance on the weak-strong cluster networks problem. QMC scales like QA and both scale better than SA. We believe this is the case because both problems are characterized by rugged energy landscapes for which tunneling transitions are a more useful way to reach low energy states than thermal transitions. In Table~\ref{NPP_Table} we also give pointers to some more efficient algorithms that achieve asymptotically better performance by exploiting existing problem structure.

 \begingroup
\squeezetable
\begin{table}
\begin{ruledtabular}
\begin{tabular}{c|c}
Method & $\alpha$\\
\hline
\specialcell{Branch and bound in conjunction with  \\ Karmarkar and Karp algorithm \cite{korf1998complete}}& 0.87\\
 \hline
 Simulated Annealing & 0.98 \\
 \hline
 Quantum Monte Carlo & 0.81 \\
 \hline
 \specialcell{Quantum Adiabatic Algorithm\\ (exact diagonalization to obtain gap scaling)} & 0.80 \\
 \hline
  \specialcell{Quantum Adiabatic Algorithm \\(time-dependent  Schr{\"o}dinger equation \cite{smelyanskiy2002dynamics}) }& 0.80 \\
  \hline
  \specialcell{Quantum Annealing\\(quantum trajectory method)} & 0.82 \\
\hline \hline
  Moduli + representations~\cite{howgrave2010new} & 0.337 \\
  \hline
  Moduli + representations + overlap~\cite{becker2011improved} & 0.291 \\
  \hline \hline
  Quantum walk + moduli + representations~\cite{bernstein2013quantum} & 0.241
\end{tabular}
\end{ruledtabular}
 \caption{Runtime scaling exponent for different methods to solve the Number Partitioning Problem. Values for the top rows are obtained from a fit of the runtime $T \propto 2^{\alpha N}$ against numerically obtained runtimes. Notice the very poor scaling of Simulated Annealing. This is because for Simulated Annealing to work at all it has to be run at very high temperatures to overcome the enormous energy barriers present in this problem. However, at these high temperatures, SA behaves almost like random sampling and hence its scaling is almost that of exhaustive search. The scaling of Quantum Monte Carlo as well as other methods that simulate Hamiltonian evolution are comparable to each other and they all scale better than SA. This mirrors the situation we encountered for the weak-strong cluster networks. However, these scalings are close enough to warrant an estimation of the standard error in these values, which we have not done. The value of $\alpha$ for the solution of the time-dependent  Schr{\"o}dinger equation was initially obtained in   \cite{smelyanskiy2002dynamics}. We also give references for the state of the art classical and quantum algorithms in the bottom rows of the table. }
 \label{NPP_Table}
\end{table}
\endgroup

To achieve such scaling behavior it will be necessary for the size  of the  domains of cotunneling qubits to grow with the problem size. However it is interesting to explore the problem from a different perspective and ask the question: {\it ``How much can the residual energy be lowered in QA  until the system reaches such a state where  lowering the energy  further  would require cotunneling of spin domains with sizes greater than $\kappa$?"}

 To answer this question one can use  an algorithm in which one starts at a random initial state and, at each step, i) looks at all groups of bits of size $\kappa$ and ii) flips  the bits  of the group that results in the largest reduction of residual energy and then one iterates. We call this procedure ``Algorithmic Tunneling" (AT). In other communities this would be referred to as $\kappa$-opt or large neighborhood search. We emphasize that AT does not provide any information about the actual system dynamics during QA, nor the runtime of QA. We investigate AT as an upper-bound on the typical performance of QA. AT does not consider the entropic component of tunneling events arising from the statistics of different mechanisms for arriving in the same minima. Likewise, AT does not consider how the height of energy barriers effects tunneling events. However, AT allows us to develop our intuition about the value of an optimal tunneling procedure which always finds the lower energy solution within a finite Hamming distance.

To investigate the lowest residual energies that AT can reach we focus on the  conditional distribution of the signed partition residues $\Omega^{\prime}$ (\ref{eq:NPP}) over  all possible spin configurations
 $\{{\bf s}^{\prime}\}$ generated from a given (ancestor) configuration ${\bf s}$ by simultaneously flipping a fixed number of spins $\kappa$. This conditional distribution  was studied in  \cite{smelyanskiy2002dynamics} and has the form
\begin{align}
P_\kappa(\Omega| \Omega') &=\frac{1}{\binom{N}{\kappa}\,P(\Omega)}\,\int_{-\infty}^{\infty}\int_{-\infty}^{\infty}\frac{dx\,dx'}{8\pi^2}\,\zeta(x+x')  \label{eq:Pr}  \\
\times & e^{\frac{i (\Omega-\Omega')x}{2}}e^{\frac{i (\Omega+\Omega')x'}{2}} \sum_{{\bf J}} \prod_{j\in {\bf J}}\cos(a_j x) \prod_{j\notin {\bf J}}\cos(a_j x') \nonumber
\end{align}
where $P(\Omega)$ is given in (\ref{eq:p}) and $\zeta(s)=\sin(\Delta \Omega s/4)/((\Delta \Omega s/4)$. In the distribution $P_\kappa(\Omega, \Omega') $, we averaged over the residue of the ancestor configuration within the interval $\Omega_{\bf s}\in [\Omega-\Delta \Omega/2,\Omega+\Delta \Omega/2]$ where $\langle E\rangle/\binom{n}{\kappa}\ll \Delta \Omega\ll \Omega$.

Near its maximum, the distribution $P_\kappa(\Omega| \Omega')$ has the form

\begin{equation}
P_{\kappa}(\Omega'|\Omega)=\frac{1}{\sqrt{2\pi
N\sigma^{2}(q)}}\exp\left[-\frac{\left(\Omega'-q\,\Omega\right)^{2}}{2N\sigma^{2}(q)}\right]\;,
\label{conditional}
\end{equation}
where 
\begin{equation}
\sigma(q)=\langle a^2\rangle\,(1-q^2)^{1/2}, \qquad q=
1-\frac{2\kappa}{N}\;.\label{sigmaq}
\end{equation}
For small $\kappa\ll N$ the width of the distribution is approximately $\sqrt{\kappa\langle a^2\rangle}$. After a single step of the AT algorithm, the average partition residue is reduced by a factor of $1-2\kappa/N$. Therefore, once the number of steps of AT  far exceeds $N/\kappa$, the algorithm reaches the residues $|\Omega | \ll  \sqrt{\kappa\langle a^2\rangle}$. For those residues the distribution becomes
\begin{equation}
P_{\kappa}(\Omega'|\Omega)\simeq P_\kappa(0|0)=\frac{1}{\sqrt{ 8\pi \kappa \langle a^2\rangle}}\;\label{eq:P0}.
\end{equation}

In total we have $\binom{N}{\kappa}$ samples of this distribution (spin configurations located on a Hamming distance $\kappa$ from the ancestor configuration). Therefore the minimum value is given by extreme order statistics \cite{galambos1980asymptotic}. I.e., within the set of bit configurations   $\{{\bf s}^{\prime}\}$ generated from a given (ancestor) configuration ${\bf s}$ by simultaneous flipping of a fixed number of spins $\kappa$, the   minimum partition energy $E$ is a random variable drawn from the exponential distribution 
\begin{equation}
p_0(E)=\frac{1}{E_\kappa} e^{-E/E_\kappa} ,\quad E_\kappa= \left [  \binom{N}{\kappa} P_\kappa(0|0) \right ]^{-1}.
\end{equation}
Therefore, average (and median) values of the minimum partition residue achieved in the AT algorithm, $E_{\rm min}^{\rm AT}(\kappa)= E_\kappa$. For $1\ll \kappa \ll N$, we obtain
\begin{equation}
E_{\rm min}^{\rm AT}(\kappa)=4\pi \kappa \langle a^2\rangle \left(\frac{\kappa}{N} \right)^{\kappa}  e^{-\kappa}. \label{eq:AT}
\end{equation}

It is instructive to compare the minimum cost values reached in AT and KK heuristics. Using (\ref{eq:KK}) and (\ref{eq:AT}) we get
\begin{equation}
\frac{E^{\rm AT}_{\rm min}}{E^{\rm KK}_{\rm min}}\sim N^{\alpha \log N -\kappa} \, \kappa^\kappa e^{-\kappa},\quad \alpha=0.72\;.\label{eq:r1}
\end{equation}
One can see from here that  tunneling over barriers of length  $\kappa >  \alpha  \log N$ allows AT to reach
cost values lower than that of the KK heuristics as $N$ increases.  

Consider AT with values of $\kappa$ that do not scale with $N$. We observe that in asymptotic limit
\begin{equation}
N\gg N_\kappa=\frac{e}{\kappa}\exp(\kappa/\alpha)\label{eq:comp},
\end{equation}
the KK heuristic produces smaller residues than AT. 
If we consider tunneling with $\kappa=8$, corresponding to the case of the weak-strong clusters studied in this paper, then $N_\kappa\simeq 22735$. We observe that for the high-precision ($B \geqslant N$) instances, NPP becomes intractable already for $N > 100$. Thus, for a broad range of problem sizes AT reaches cost values much smaller than conventional heuristics.

 Motivated by the above observation we conclude that asymptotic scaling behavior is not essential for this analysis. For example, one can choose AT with the barriers sizes  $\kappa=\alpha_0 \log N$ where $\alpha_0-\alpha \gg 1/\log N$. In this case the ratio
$E^{\rm AT}_{\rm min}/E^{\rm KK}_{\rm min}$ approaches 0 in the asymptotic limit  $N\rightarrow\infty$. 
However the length of the barriers remains relatively small in a very broad range of $N$.  For example, consider   $\alpha_0=1.16$.
Then for $N=1000$ the barrier  length is $\kappa=10$  while $E^{\rm AT}_{\rm min}/E^{\rm KK}_{\rm min}\sim 10^{-9}$.

In summary, a greedy search procedure with flipping at most $\kappa$ bits at a time (referred above as Algorithmic Tunneling) allows to find cost values in NPP that are much below those given by KK heuristics for $\kappa\sim10$ at all realistic values of $N$. It would be interesting to compare the minimum residue obtained by AT with the minimum residue obtained by the other algorithms in Table~\ref{NPP_Table} when constrained to terminate in polynomial time.

\subsection{$K^\textrm{th}$ Order Binary Optimization}

Our current best candidate for a problem class that fulfills all three
criteria consists of $K^\textrm{th}$ order binary optimization problems with $K>2$. $K^\textrm{th}$
order binary optimization is NP-Hard and occurs naturally in many engineering
disciplines and many computational tasks. In
unpublished work underway, we seek to establish that for many
$K$-local problems, QA indeed offers a runtime advantage over SA. Currently we are focusing on $K \in \{ 4,5,6 \} $. As energy
landscapes get more rugged with higher $K$, our conjecture is that we
will see larger subsets of instances for which QA runs faster as $K$
increases. However, representing $K$-body terms in hardware becomes more
challenging as $K$ grows.

\section{Designing Future Annealers of Practical Relevance}\label{kspin_challenges}

Should numerical studies confirm that QA offers a substantial runtime advantage, there is still one more hurdle to overcome. We need to ensure that $K$-local problems can be economically represented in hardware. We would like to be able to tell a user: ``If you have a binary optimization problem with $N$ variables and $L$ terms, and the many-body order of the highest term is $K$, then you can send this problem to the quantum annealing co-processor.'' However, annealers built to-date only support pairwise qubit couplings, i.e. $K=2$. Two routes have been proposed to increase the locality. 

One route is to build physical $K$-body couplers. However, it may
prove difficult to layout $K$-local couplers on a two dimensional chip
or even in layered architectures. Of course, the
general case in which one aims to implement all possible $
\sum_{k=1}^K {N\choose k}$ couplings will be infeasible. While many
applications will only necessitate $L=\mathcal O(N)$ coupling terms, this could
still prove challenging. Furthermore, economically embedding problems in a fixed
graph with only a limited number of specific $K$-local terms may prove
difficult.

Another possibility is that we could use logical embeddings that map $K$-local problems
to 2-local problems. A new proposal on how to accomplish
such embeddings has been put forth~\cite{lechner_quantum_2015}, which has reinvigorated interest in this direction. Our main worry regarding any reduction to quadratic problems is that this will involve ancillary qubits. As argued above, it is crucial that the problem features tall and narrow barriers for tunneling transitions to contribute and the introduction of additional variables may cause these barriers to become wider. This makes purely thermal annealing more competitive and may negate gains seen in the numerics prior to embedding. 

\section{Summary}\label{summary}

It is often quipped that Simulated Annealing is only for the ``ignorant or desperate''.
Yet, in our experience we find that lean stochastic local search
techniques such as SA are often very competitive and they continue to
be one of the most widely used optimization schemes. This is because
for sufficiently complex optimization tasks with little structure to
exploit (such as instances of $K^\textrm{th}$  order binary optimization) it often
takes considerable expert effort to devise a faster method. Therefore we regard SA as the generic classical competition that Quantum Annealing needs to beat. 

Here we showed that for carefully crafted proof-of-principle problems with rugged energy landscapes that are dominated by large and tall barriers QA can have a significant runtime advantage over SA. We found that for problem sizes involving nearly 1000 binary variables, quantum annealing is more than $10^8$ times faster than SA running on a single core. We also compared the quantum hardware to QMC. While the scaling of runtimes with size between these two methods is comparable, they are again separated by a large factor sometimes as high as $10^8$.

For higher order optimization problems, rugged energy landscapes will become typical. As we saw in our experiments with the D-Wave 2X, problems with such landscapes stand to benefit from quantum optimization because quantum tunneling makes it easier to traverse tall and narrow energy barriers. Therefore, we expect that quantum annealing might also deliver runtime advantages for problems of practical interest such as $K^\textrm{th}$ order binary optimization with larger $K$.

More work is needed to turn quantum enhanced optimization into a practical technology. The design of next generation annealers must facilitate the embedding of problems of practical relevance. For instance, we would like to increase the density and control precision of the connections between the qubits as well as their coherence. Another enhancement we wish to engineer is to support the representation not only of quadratic optimization but of higher order optimization as well. This necessitates that not only pairs of qubits can interact directly but also larger sets of qubits. Such improvements will also make it easier for end users to input hard optimization problems. 

The work presented here focused on the computational resource that is experimentally most accessible for quantum annealers: finite range tunneling. However this analysis is far from complete. A coherent annealer could accelerate the transition through saddle points, an issue slowing down the training of deep neural networks, for reasons similar to those that make a quantum walk spread faster than a classical random walker~\cite{ambainis2003quantum,childs2003exponential,kempe2003quantum}. It could also dramatically accelerate sampling from a probability distribution via the mechanism of many body delocalization~\cite{baldwin_many-body_2015}. The computational value of such physics still needs to be properly understood and modeled.

\vspace{0.5in}
\textbf{Acknowledgements --} We would like to thank Matthias Troyer for discussions and Edward Farhi and Masoud Mohseni for helping to review the manuscript. We would also like to thank Daniel Lidar, Damian Steiger, Alex Selby, and Dvir Kafri for comments. 

\appendix

\section{Quantum Annealing  results for Weak-Strong cluster problem with 16 qubits}\label{app:modelling}

We developed a detailed modeling of the quantum annealing
process and incoherent multi-qubit cotunneling for the weak-strong
cluster problem in Ref.~\cite{boixo_computational_2014}. Using a noise
model with experimentally measured parameters for the D-Wave 2X
processor, we numerically verified that the spins arrive at the
energetically more favorable configuration via multi-qubit
tunneling. In what following we will refer to the modeling in this
reference for the details.

 \begin{figure}[tb]
 \includegraphics[width=1.0\columnwidth]{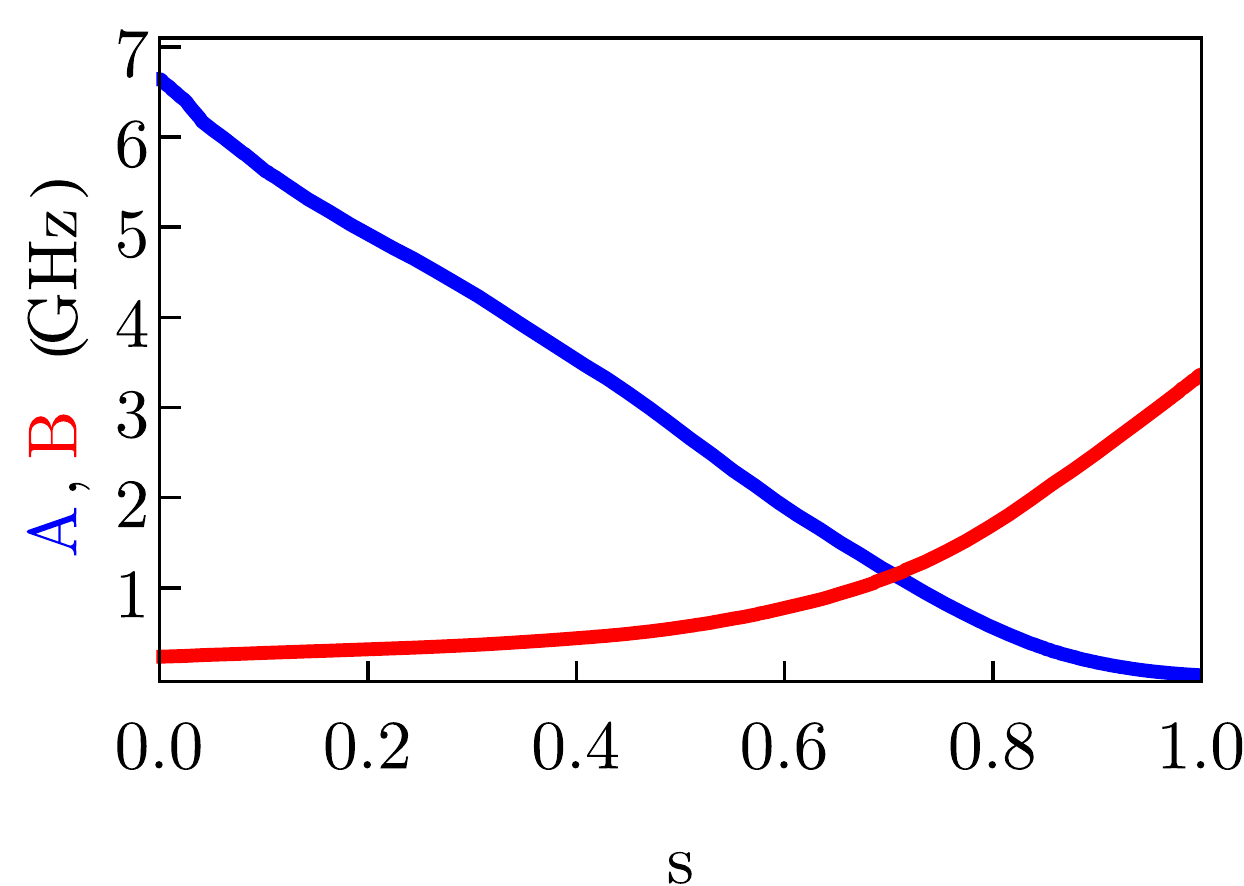}
 \caption{ Schedule functions for D-Wave 2X chip. The annealing parameter is $s= t/T_{\rm QA}$ for time $t$ and total annealing time $T_{\rm QA}$.
 }
\label{fig:AB}
 \end{figure}

In the present study we apply this detailed model for the new schedule
functions $A(s)$, $B(s)$ (see Fig.~\ref{fig:AB}) and for the new
values of the noise parameters, line-width $W$, Ohmic coefficient
$\eta$, and temperature of the device $T$ for the D-Wave 2X processor.
The noise parameters were measured near the end of the quantum
annealing schedule, $s=1$.  The values of the noise parameters at a
point during the annealing can be related to the measured ones (see
Appendix A 5 in~\cite{boixo_computational_2014})
\begin{align}
&(W(s)/W_{\rm MRT})^2=\eta/\eta_{\rm mRT}=B(s)/B(1) \label{eq:par}\\
&W_{\rm MRT}\simeq 661\, {\rm MHz},\quad \eta_{\rm MRT}\simeq 0.12,\quad T \simeq 12\, {\rm mK}\;.\nonumber
\end{align}

\begin{figure}[h]
  \centering
  \includegraphics[width=\columnwidth]{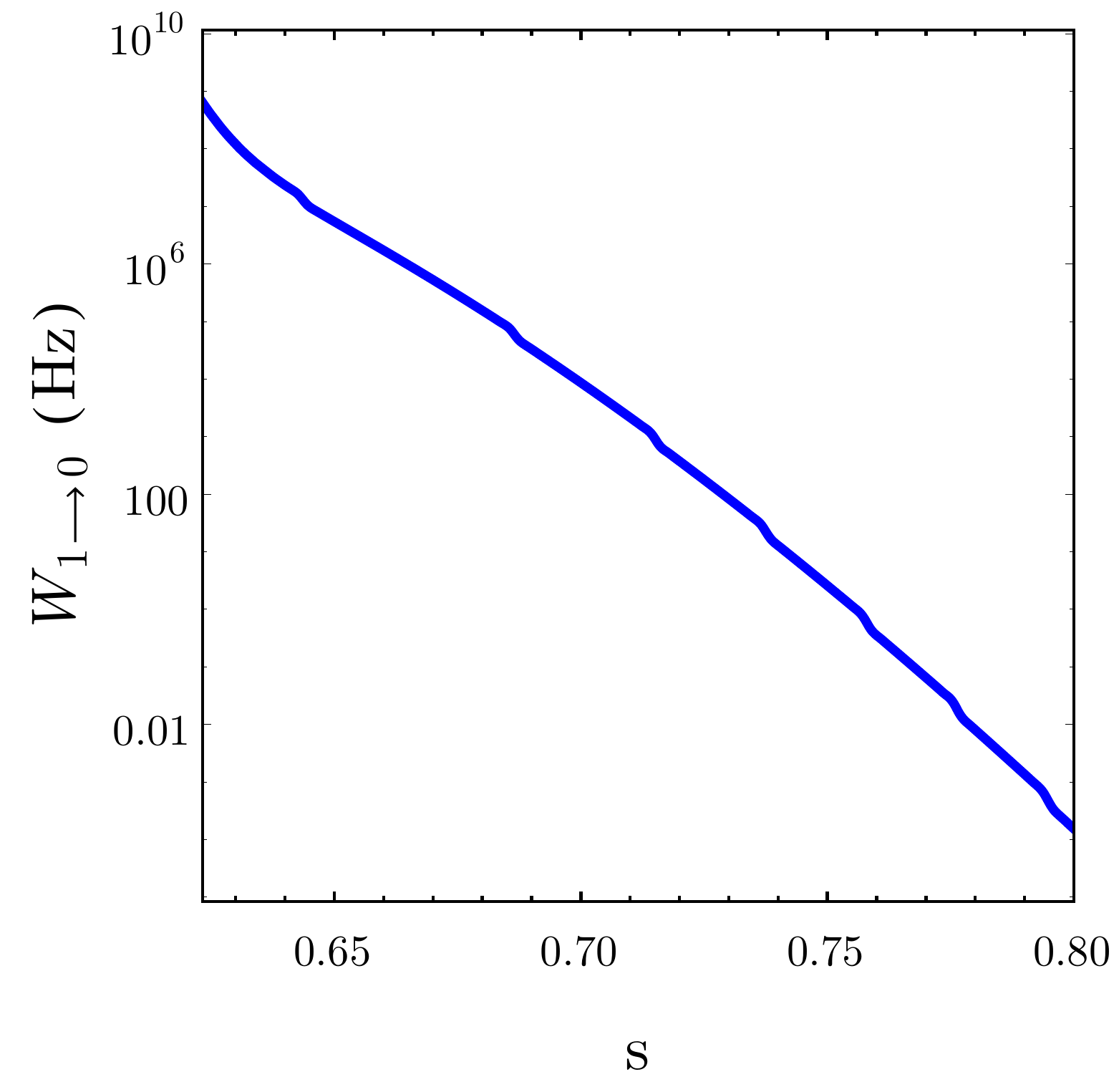}
  \caption{Logarithmic plot of the transition rate $W_{10}(s)$ vs $s$ after the avoided crossing. Plot corresponds to $T=12mK$ and $h_1=0.44$.}
  \label{fig:W10}
\end{figure}

\begin{figure}[h]
  \centering
  \includegraphics[width=\columnwidth]{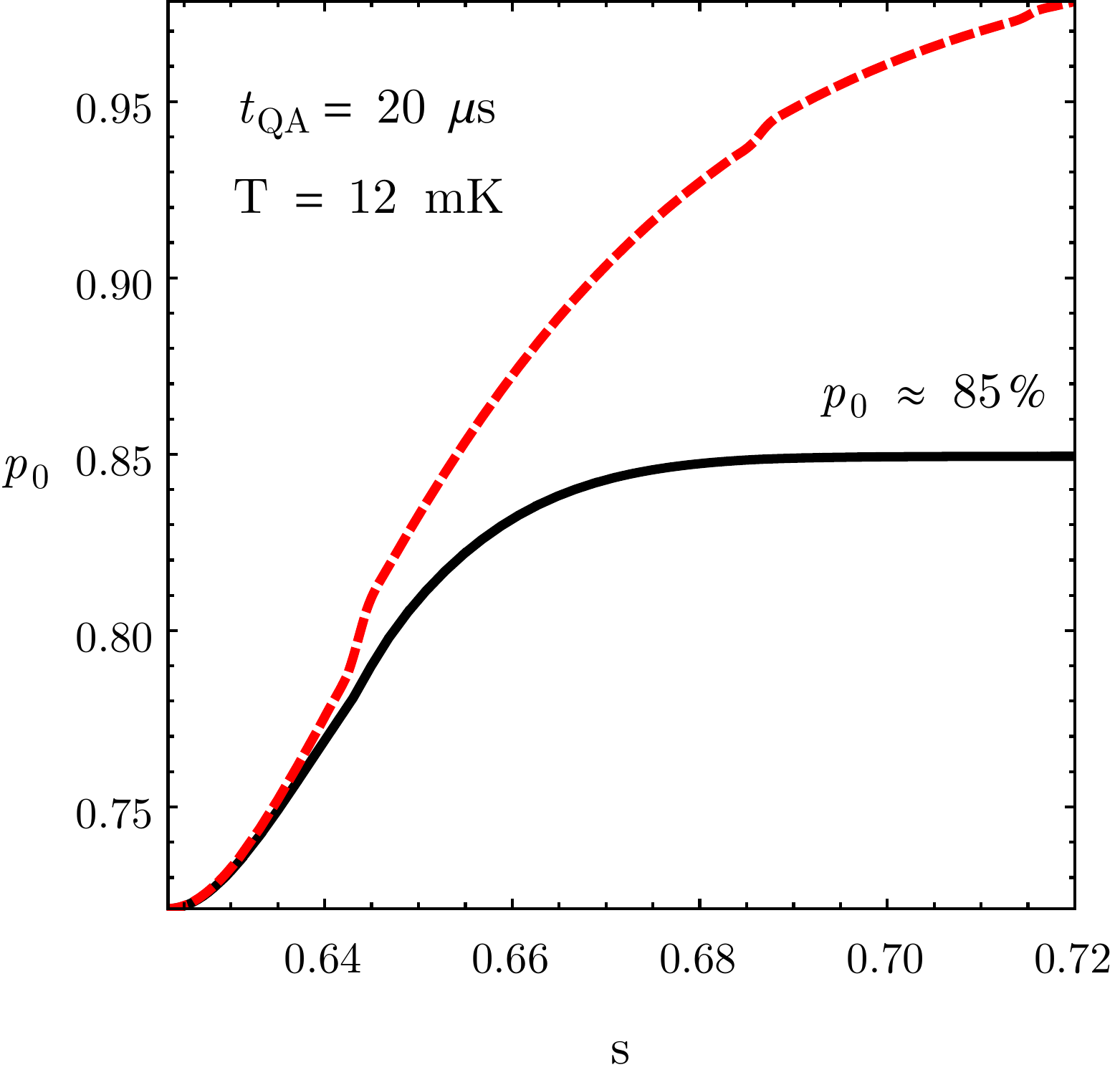}
  \caption{Probability of occupation of the ground state during the QA
    process is shown with solid black line. Red dashed line gives the
    equilibrium population of the ground state. Results are given at
    $T=12 {\rm mK}$ and $h_1=0.44$. Dashed red line gives the
    equilibrium population of the ground state. The quantum annealing
    time is $t_{\rm QA}$=20 $\mu$s. }
  \label{fig:p0}
\end{figure}

 The population $p_0(t)$ of the ground state during the QA process obeys the equations
\begin{align}
&\frac{d p_0}{dt} =-(W_{01}(s)+W_{10}(s)) p_0(s)+W_{10}(s) ,\label{eq:p0}\\
&\frac{W_{01}(s)}{W_{10}(s)} =e^{-\Delta_{10}(s)/k_B T},\quad \Delta_{10}(s)=E_1(s)-E_0(s)\;,\nonumber
\end{align}
where $W_{jk}(s)$ is a transition rate from the state $j$ to the state
$k$ whose explicit form is given in
Ref.~\cite{boixo_computational_2014}.  The transition rate $W_{10}(s)$
decays very fast after the avoided crossing (see Fig.~\ref{fig:W10})
because the weak cluster (left cluster in Fig.~\ref{fig:Weak_Strong})
becomes progressively more polarized along the z-axis and the
effective size of the tunneling domain $D=D(s)$ grows. This gives rise
to the multi-qubit ``freezing phenomenon'' where the system population
gets partially trapped in the excited state after certain value of $s$
in later stages of QA ~\cite{boixo_computational_2014}. Fig.~\ref{fig:p0} shows the 
ground state population given by the solution of (\ref{eq:p0}).  The
success probability to be at the ground state at the end of the
annealing is $p_0=0.85$, which is close to the experimentally
observed mean value of 0.9. It is seen that the equilibrium population of
the ground state exceeds the actual population for $s\gtrsim 0.64$,
corresponding to the onset of freezing of the transition rates.

\begin{figure}[h]
  \centering
  \includegraphics[width=3.2in]{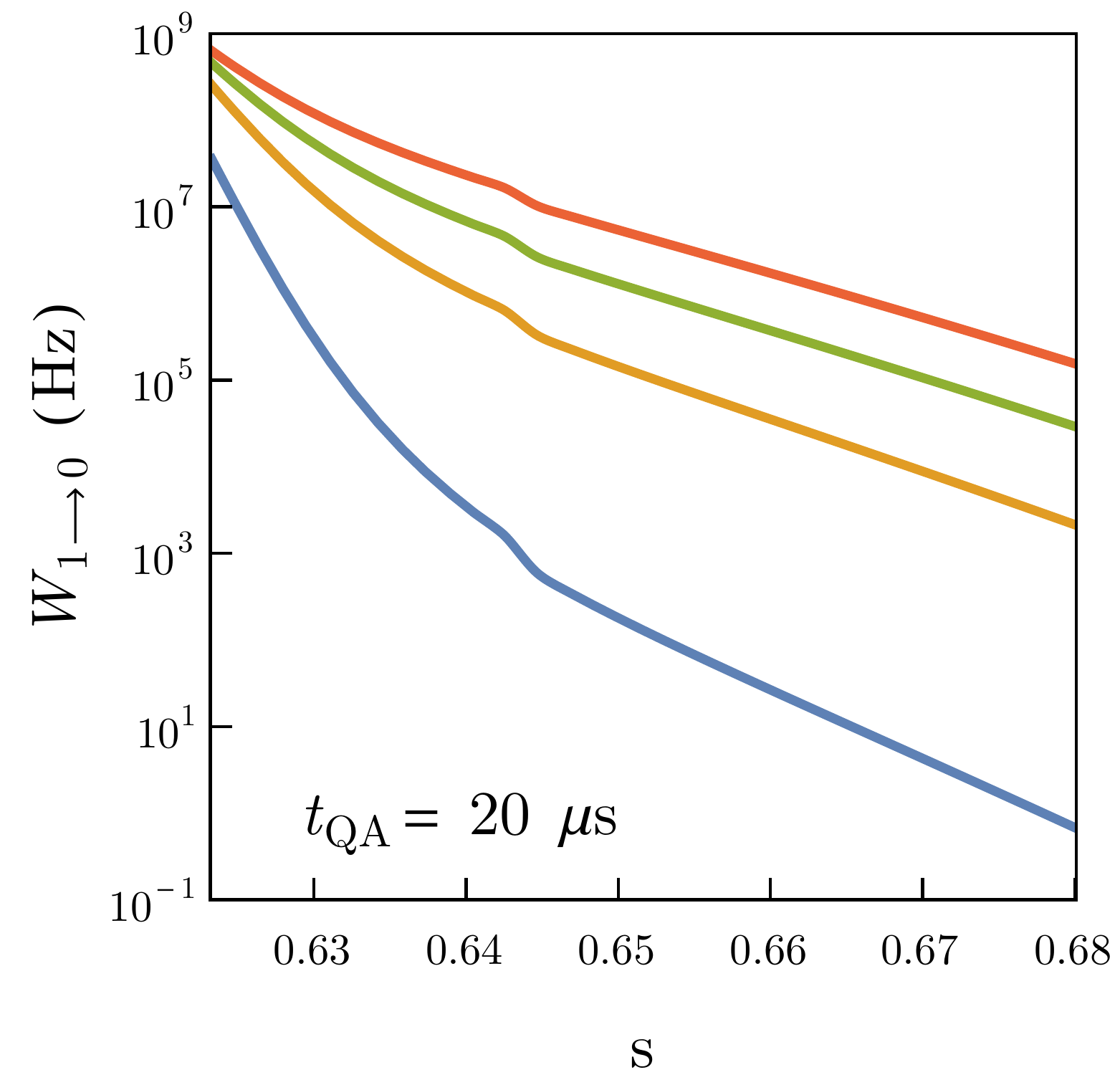}
  \caption{Logarithmic plots of the transition rate $W_{10}(s)$ vs $s$ after the avoided crossing for different temperatures. Red, green, mustard and blue colors corresponds to $T$ = 12mK, 8 mK, 6 mK and 4 mK respectively. All plots correspond to $h_1=0.44$ and $T_{\rm QA}$=20$\mu$sec.}
  \label{fig:W10T}
\end{figure}

\begin{figure}[h]
  \centering
  \includegraphics[width=\columnwidth]{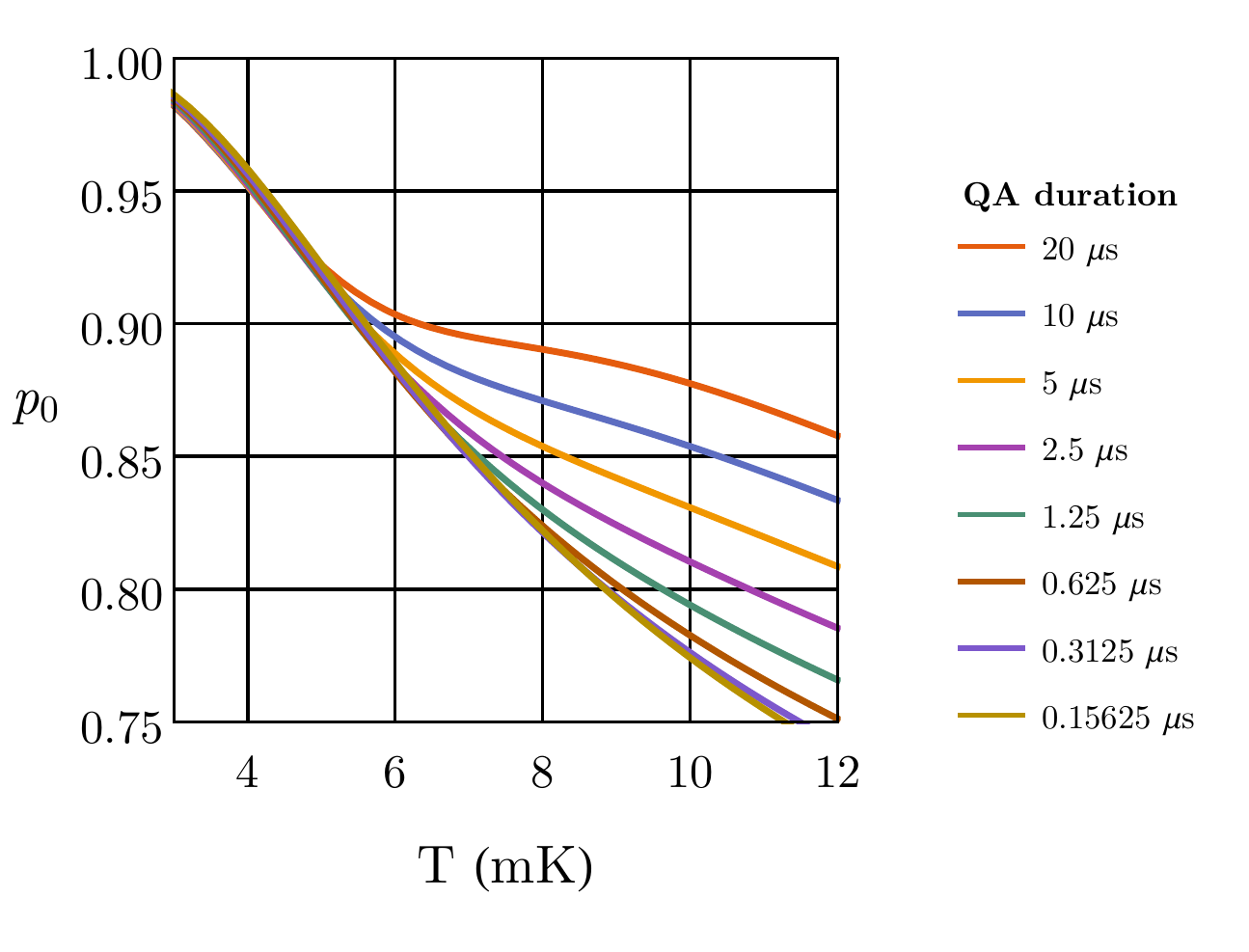}
  \caption{Success probability of QA $p_0$ vs $T$. Different colors correspond to different durations of the QA process $T_{\rm QA}$. Plots correspond to  $h_1=0.44$.}
  \label{fig:p0T}
\end{figure}

\begin{figure}[h]
  \centering
  \includegraphics[width=\columnwidth]{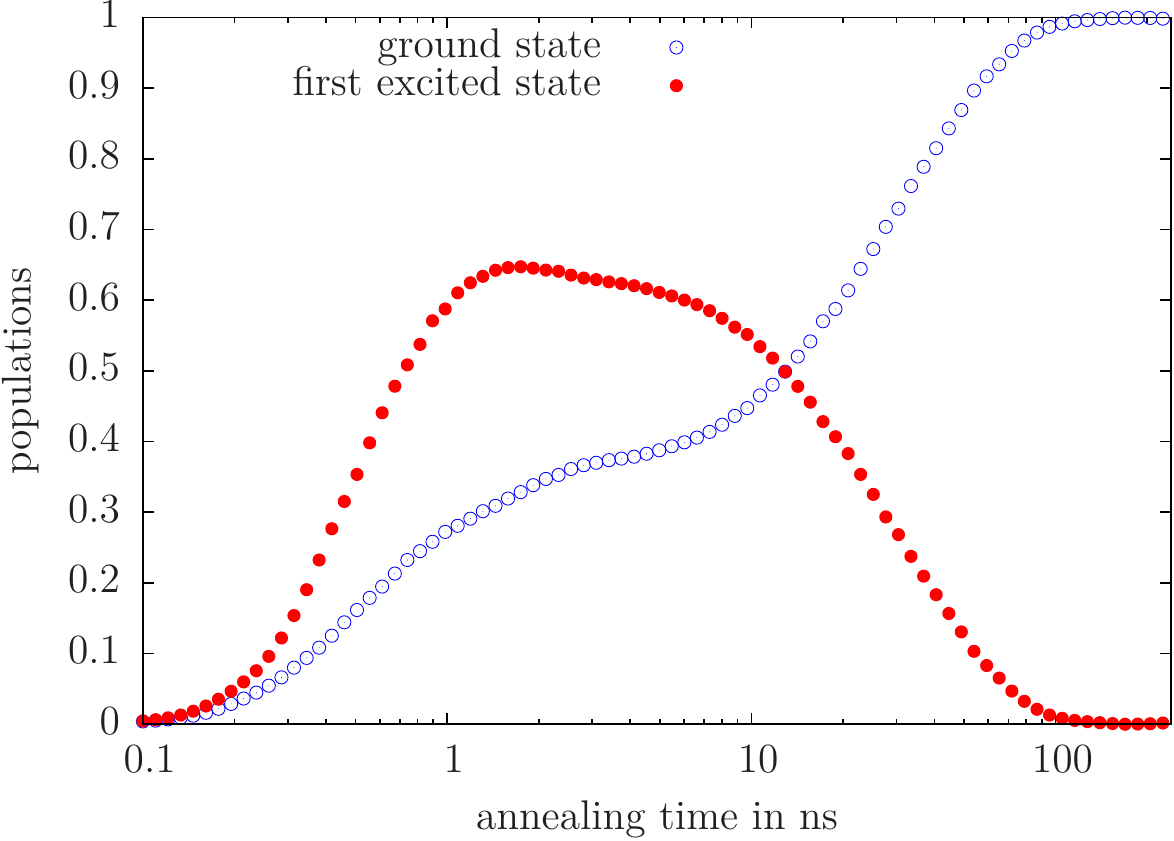}
  \caption{Plots show the results of the solution  of the time-dependent Schr\"odinger equation for the closed system QA. Probabilities of occupation of ground state and first excited state as a function of QA time are plotted with blue and red points, respectively. The QA time to reach the probability of success 0.95 equals to  70.9 ns.
   All plots correspond to  $h_1=0.44$.}
  \label{fig:TDS}
\end{figure}

\section{D-Wave versus Quantum Monte Carlo with Linear Schedule}\label{app_b}

\begin{figure}[h!]
  \centering
  {\includegraphics[width=\columnwidth]{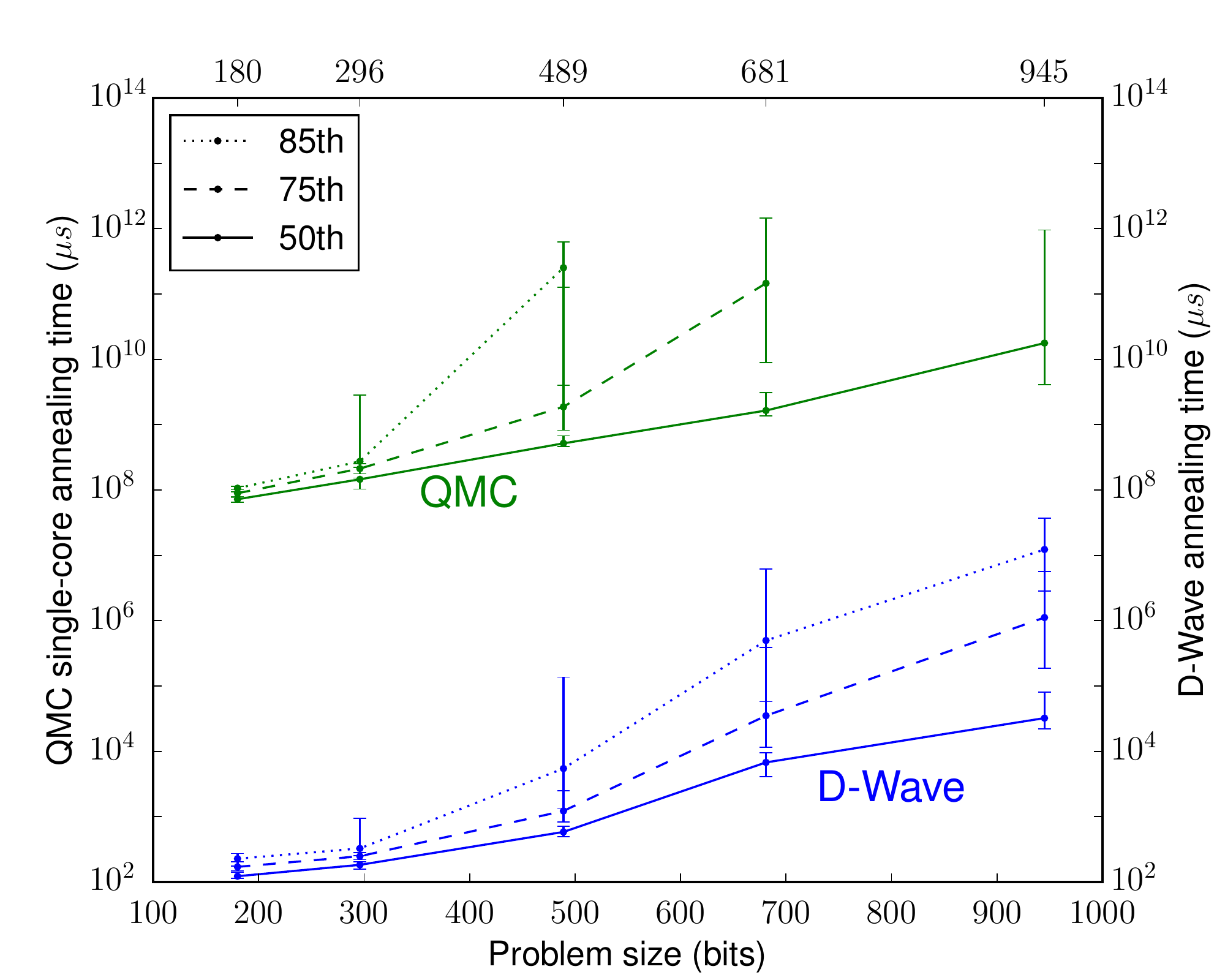}}
  \caption{Time to find the optimal solution with 99\% probability for different problem sizes. We compare Quantum Monte Carlo (QMC) with a linear schedule and the D-Wave 2X. To assign a runtime for QMC we take the number of worldline updates that are required to reach a 99\% success probability and multiply that with the time to perform one update on a single state-of-the-art core. Shown are the 50th, 75th and 85th percentiles over a set of 100 instances. The error bars represent $95\%$ confidence intervals from bootstrapping. The runtimes for the higher quantiles for the larger problem sizes for QMC were not computed because the computational cost was too high.}
\label{QMC_linear_versus-D-Wave_Timings}
\end{figure}

There is ongoing work directed to optimize the QMC parameters further. In preliminary results, we compared QMC with a linear schedule against D-Wave~\footnote{The annealing function $A(s)$ decreases linearly from 1 to 0, while $B(s)$ increases linearly from 0 to 1.}. The transverse field in this case is lower, resulting in a faster time $T_{\rm worldline}$ to update a worldline ($T_{\rm worldline}$ scales linearly with the transverse field).  We measured this time to be 
\begin{equation}
T_{\rm worldline}= \beta \times 115 \;{\rm ns} \;.\label{eq:Twl}
\end{equation} 
This time is consistent with the one reported in Ref.~\cite{boixo_evidence_2014}. 

We also took a different approach when optimizing $\beta$ and the number of sweeps per run to minimize the total computational effort. In the case reported in Sec.~\ref{sec:d-wave_vs_qmc} we optimized the number of sweeps for each quantile at fixed $\beta = 10$. In the case of a linear schedule, we used our knowledge of the structure of the weak-strong cluster networks problem to optimize $\beta$ and the number of sweeps $n_{\rm sweeps}$ concurrently. We first measure the probability of success $p(n_{\rm sweeps}, \beta)$ for a single weak-strong cluster pair. Then we estimate the performance for the cluster network problems taking into account the number of cluster pairs $c$ for each size. The estimate is
\begin{align}
  \textrm{total time} \propto n_{\rm sweeps} \times \beta \times \left\lceil \frac {\log(1-0.99)}{\log(1-p(n_{\rm sweeps}, \beta)^c)} \right\rceil  \;.\nonumber
\end{align}
Here we also run with open boundary conditions (OBC) in imaginary time. We estimate an optimal $\beta = 130$ for all sizes. We then optimize the number of sweeps for each quantile and size running the actual benchmark. Finally, we modified the path integral QMC code to search for the minimum energy configuration along all replicas at the end of the annealing.

The results, following the same methodology as in Sec.~\ref{sec:d-wave_vs_qmc}, are plotted in Fig.~\ref{QMC_linear_versus-D-Wave_Timings}. We obtain a prefactor $\sim 10^6$ for the median and up to $\sim 10^8$ for the 85th quantile. 

When optimizing QMC to the extent performed here a methodological concern arises. Since QMC has many parameters and modes of execution (e.g. temperature, number of sweeps, annealing schedule, open versus closed boundary conditions in imaginary time, discrete or continuous time Monte Carlo), overlearning can become an issue when working with just 100 instances. Moreover, optimizations over many parameters will become computationally prohibitive as the problem size increases. By contrast, the current quantum hardware has only a single parameter that can be tuned, the number of annealing sweeps.

\clearpage

\bibliography{Finite_Tunneling}

\end{document}